\documentclass[sn-mathphys,iicol]{sn-jnl-arxiv}
\usepackage{amsmath}
\usepackage{ar}

\DeclareMathOperator{\pe}{\mathit{Pe}}
\DeclareMathOperator{\Rey}{\mathit{Re}}
\DeclareMathOperator{\sch}{\mathit{Sc}}
\DeclareMathOperator{\pr}{\mathit{Pr}}

\DeclareMathOperator{\raT}{\!\mathit{Ra}_T}
\DeclareMathOperator{\raM}{\!\mathit{Ra}}
\DeclareMathOperator{\raD}{\!\mathit{Ra}_\mathit{d}}

\DeclareMathOperator{\da}{\mathit{Da}}
\DeclareMathOperator{\sh}{\mathit{Sh}}

\jyear{2023}

\begin{document}

\title{Convective mixing in porous media: A review of Darcy, pore-scale and Hele-Shaw studies}

\author*[1,2]{\fnm{Marco} \sur{De Paoli}}\email{m.depaoli@utwente.nl}

\affil[1]{\orgdiv{Physics of Fluids Group}, \orgname{University of Twente}, \orgaddress{\street{P.O. Box 217}, \city{Enschede}, \postcode{7500AE}, \country{the Netherlands}}}

\affil[2]{\orgdiv{Institute of Fluid Mechanics and Heat Transfer}, \orgname{TU Wien}, \orgaddress{\street{Getreidemarkt 9}, \city{Vienna}, \postcode{1060}, \country{Austria}}}

\abstract{
Convection-driven porous media flows are common in industrial processes and in nature. The multiscale and multiphase character of these systems and the inherent non-linear flow dynamics make convection in porous media a complex phenomenon. As a result, a combination of different complementary approaches, namely theory, simulations and experiments, have been deployed to elucidate the intricate physics of convection in porous media. In this work, we review recent findings on mixing in fluid-saturated porous media convection. We focus on the dissolution of a heavy fluid layer into a lighter one, and we consider different flow configurations. We present Darcy, pore-scale and Hele-Shaw investigations inspired by geophysical processes. While the results obtained for Darcy flows match the dissolution behaviour predicted theoretically, Hele-Shaw and pore-scale investigations reveal a different and tangled scenario in which finite-size effects play a key role. Finally, we present recent numerical and experimental developments and we highlight possible future research directions. The findings reviewed in this work will be crucial to make reliable predictions about the long-term behaviour of dissolution and mixing in engineering and natural processes, which are required to tackle societal challenges such as climate change mitigation and energy transition.}



\maketitle

\tableofcontents

\section[Introduction]{Introduction}\label{sec:intro}
\begin{figure*}
\center
\includegraphics[width=0.98\textwidth]{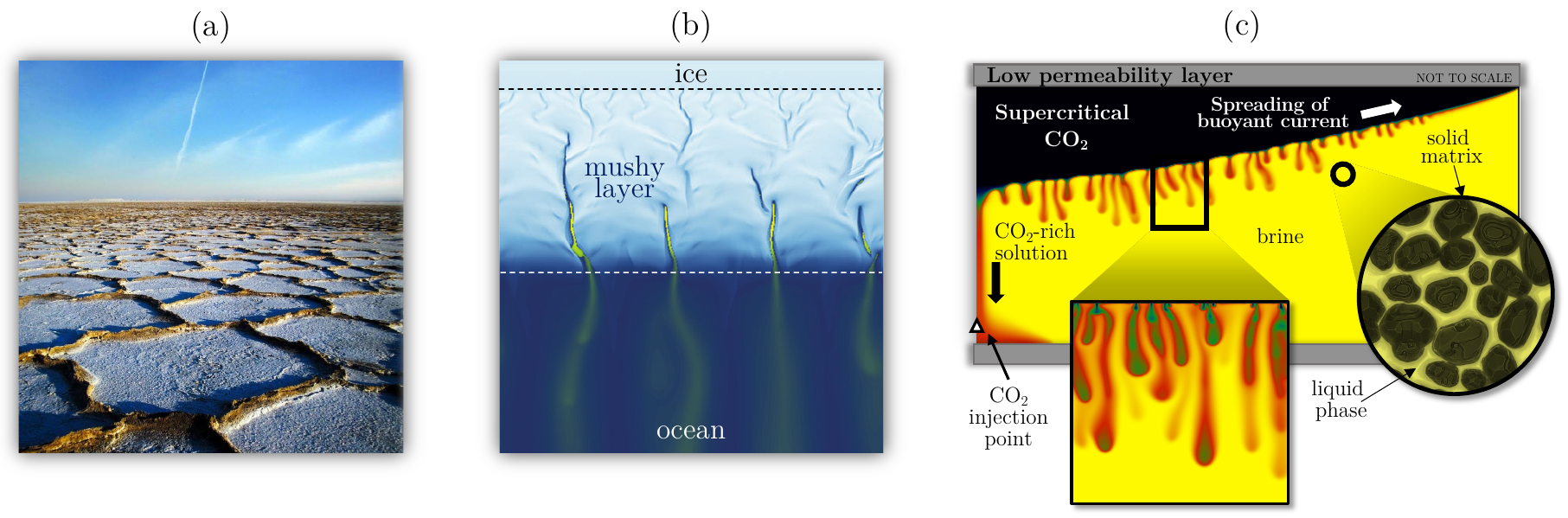}
\caption{\label{fig:intro}
Examples of convection in porous media in geophysical applications.
(a)~Salt polygons at the Hoz-e Soltan (Iran) \cite[image courtesy of][]{picturesalt}.
These superficial formations are the result of salt-induced convective subsurface flows. 
(b)~Formation of sea ice \cite[adapted with permission from][]{anderson2022mushy}. 
When sea ice grows, the intermediate layer between the ice exposed to the atmosphere and the ocean forms a porous solid matrix (ice) filled in the interstitial space by brine (water and salt).
Salt-rich (yellow) plumes of brine drain from this mushy layer into the underlying ocean (blue).
(c)~Migration of carbon dioxide (CO$_2$) in a post-injection scenario \cite[adapted with permission from][]{depaoli2021influence}.
Brine and CO$_2$ saturate the porous medium and are vertically confined by two low-permeability layers.
Due to symmetry, only the right half of the reservoir is shown. (square)~Dissolution of CO$_2$ in brine occurs at the interface between the currents of these fluids. (circle)~Liquid phase filling the interstitial space within the pores of the rocks.
}
\end{figure*}

A porous medium is a material consisting of a solid matrix with an interconnected void, which allows fluids to flow through it.
When a fluid-saturated medium subject to the action of gravity experiences an unstable density profile, i.e. a heavy fluid parcel sitting above a less dense one, the denser fluid will eventually move and replace the lighter fluid, and vice versa.
The density-driven physical mechanism inducing this motion is defined as convection, and it represents the driving force of many problems of practical interest, particularly in geophysical processes.
The regular polygonally patterned crusts of salt shown in Fig.~\ref{fig:intro}(a), approximately a meter in diameter, are the surface signature of the vertical transport of salt, a fundamental process in arid regions.
These ridges form as a result of solutal convection in the porous soil beneath the surface \citep{lasser2021stability,lasser2023salt}.
Similarly, in supercritical geothermal systems heat supplied by a magmatic heat source produces a buoyancy-induced flow circulation due to convection \citep{parisio2019risks}.
Formation of sea ice (Fig.~\ref{fig:intro}b) or solidification of multicomponent alloys may originate mushy layers, which consist of a porous medium filled with interstitial fluid \cite{anderson2020convective}.
This fluid (brine, a mixture of water and sea salt) experiences density gradients produced by differences of temperature and solute concentration, which induce convective motions within the porous layer and control the subsequent solidification dynamics \cite{wells2019mushy,du2023sea}.
The above convective processes in porous media are associated with grand societal challenges, including energy transition and climate change mitigation.
Understanding the underlying fluid mechanics is crucial for making reliable predictions on the evolution of the natural environment \cite{dauxois2021confronting}.
Within the many applications of importance in this context, convection in porous media has received renovated attention due to the implications it bears for geological sequestration of carbon dioxide (CO$_{2}$) \cite{riaz2014carbon}.

Geological CO$_{2}$ storage consists of injecting large volumes of carbon dioxide in underground geological formations with the aim of permanent (or long-term) storage (Fig.~\ref{fig:intro}c). 
These formations are typically saline aquifers and consist of a porous material confined by horizontal low permeability layers (grey regions in Fig.~\ref{fig:intro}c).
The aquifers are located 1--3~km beneath the Earth surface, where the pressure is sufficient to keep the CO$_{2}$ supercritical \cite{huppert2014fluid,emami2015convective}.
Here, a rich flow dynamics emerges. 
Injected CO$_{2}$ (black) is initially lighter than the fluid (brine, yellow) naturally filling the subsurface aquifer, and therefore carbon dioxide migrates towards the upper region of the formation, driven by convection, to form a CO$_{2}$ layer that spreads horizontally.
The low permeability layer prevents CO$_{2}$ from escaping and migrating to the uppermost parts of the aquifer, from where it could eventually return to the atmosphere.
At the interface between the currents of carbon dioxide and brine, the dissolution of CO$_{2}$ into the underlying brine layer takes place, originating a new mixture heavier than both starting fluids (red-to-green fluid in Fig.~\ref{fig:intro}c).
The dissolution process, illustrated in the squared inset of Fig.~\ref{fig:intro}(c), makes the interfacial layer heavier and thicker, and eventually finger-like instabilities form.
The CO$_{2}$-rich solution will sink down being permanently stored in the formation. 
The presence of these finger-like structures makes the convective dissolution process more efficient compared to a diffusive dissolution.
Such a behaviour is highly desired for CO$_{2}$ storage because injected carbon dioxide will dissolve faster preventing leakages in case of faults at the top low-permeability confining layer.
In turn, the presence of non-linear structures makes the system complex to study, and long-term predictions of the dynamics of injected carbon dioxide require huge computational efforts.
An element further increasing the complexity of this scenario is represented by the finite-size pore-scale effects.
At the level of the rock grains, schematically reported in the circle of Fig.~\ref{fig:intro}(c), the fluid moves in the interstitial space following sinuous paths, further spreading the solute transported and making predictions on the long-term behaviour even more challenging. 
Motivated by the CO$_{2}$ storage process, convection in porous media has been recently investigated in great detail \cite{hewitt2020vigorous}. 
In this work, we will review the current modelling approaches, numerical and laboratory measurements, and in particular we will focus on the role of finite-size effects such as confinements and pore-scale dispersion.

In a convective porous medium flow, the dynamics is controlled by the relative importance of driving and dissipative mechanisms, which is quantified by the Rayleigh-Darcy number $\raM$. 
Convection is the driving process, and it is determined by the combination of fluid properties (density contrast), medium properties (permeability and porosity) and domain properties (gravity and domain size).
Dissipative forces act against convection either as a drag force between the fluid and the solid (due to viscosity) or reducing local gradients of density (due to molecular diffusion).
As a result of solute redistribution due to the tortuous fluid path in the interstitial matrix, the solute concentration field is made more uniform.
This effect, labelled as dispersion, contributes as well to dissipate the potential mixing energy of the system, since the concentration gradients within the domain reduce. 
A key challenge in studying convective geophysical flows consists of making reliable predictions of their evolution by determining how global transport quantities, e.g. the solute flux or the mixing rate, vary as a function of $\raM$ and time.
Simplified mathematical models solved numerically and theoretically have provided a clear picture of the flow behaviour at the Darcy scale \cite{pau2010high,hewitt2013convective,slim2014solutal}, i.e. when a sufficiently large representative elementary volume including many pores is considered \cite{nield17}.
In contrast, these results disagree with the experimental measurements \cite{neufeld2010convective,backhaus2011convective}, possibly suggesting that physical effects present in laboratory setups are not captured by the classical Darcy formulation \cite{hidalgo2012scaling}.

An intuitive way to experimentally mimic a porous medium consists of filling a confined volume with solid materials, and when spherical objects are used the medium is defined as a bead pack \cite{woods2015flow}. 
These experiments may be challenging, since the medium is typically hardly accessible due to its opacity, and only in recent years non-invasive and non-intrusive measurements such as X-ray tomography and magnetic resonance imaging have become accessible \cite{poelma2020measurement,tropea2007springer}.
As a result, most of the experiments on convective flows in porous media have been performed in Hele-Shaw cells, which consist of two transparent plates separated by a narrow gap where the fluid flows \cite{depaoli2020jfm}. 
The Hele-Shaw apparatus is particularly relevant because it provides optical access, and in some conditions the flow follows a Darcy-like behaviour.
In general, neither bead packs nor Hele-Shaw cells faithfully reproduce the dynamics of a Darcy flow, in which the flow structures within a porous medium are much larger than the average pore size.
A difference emerged in the transport properties of bead packs experiments, Hele-Shaw experiments and numerical simulations \cite{hidalgo2012scaling}.
The solute redistribution effects (dispersion) produced either by the presence of the solid obstacles in the porous matrix or by the walls in a Hele-Shaw flow have been identified as the main responsible for these discrepancies \cite{liang2018effect}, and are labelled here as \emph{finite-size} effects. 
In recent years, the advancement of theoretical, experimental and numerical techniques allowed a more precise characterisation of the flow, with accurate measurements of pore-scale dissolution rates, and a clearer picture of the influence of finite-size effects on dissolution and mixing is now available.

In this work, we review recent theoretical, numerical and experimental findings in the field of convection in porous media. 
This review is meant to be complementary with respect to other works \cite{hewitt2020vigorous,huppert2014fluid,emami2015convective}, since we focus on dissolution and mixing with emphasis on finite-size effects.
The paper is organised as follows. 
In Sec.~\ref{sec:conv} we describe the mathematical models and the idealized configurations used to investigate convection in porous media, and we derive a unified formulation to evaluate and relate mixing in different flow configurations.
In Secs.~\ref{sec:res2} and~\ref{sec:os}, we review the results obtained in Rayleigh-B\'enard and one-sided configurations, respectively.
Finite-size effects possibly leading to the discrepancy observed between experiments and simulations are discussed in Sec.~\ref{sec:osdisp}.
Finally, in Sec.~\ref{sec:concl} we summarise the results discussed and present recent experimental developments, together with an overview of additional effects not present in the configurations discussed in Secs.~\ref{sec:res2} and~\ref{sec:os}.

\section[Modelling of convection]{Modelling of convection}\label{sec:conv}
\subsection{Pore-scale modelling}\label{sec:convpore}
Convective flows are produced by the presence of unstable density gradients within an accelerated fluid domain.
These density differences drive the flow towards a more stable configuration, decreasing the gravitational potential energy within the system \cite{boffetta2017incompressible}.
We consider problems in which convection is induced by the presence of a scalar quantity (e.g., solute concentration or temperature) that modifies the density field of the flow. 
For simplicity, in this review we will define the parameters in case of solute convection, but the findings extend to the case of thermally-driven convection unless explicitly mentioned.

The maximum density difference within the domain, $\Delta\rho$, determines the strength of the convective flow.
On the other hand, (molecular or thermal) diffusion reduces the local scalar gradients diminishing the driving force of the flow, and viscosity is responsible for energy dissipation due to friction.
In a free fluid (i.e., in the absence of a porous medium) the relative importance of these two contributions is quantified by the Rayleigh number $\raT$ defined on the characteristic length scale of the flow $H$
\begin{align}
\raT=\frac{g\Delta\rho H^3}{\mu D}\text{  ,}
\label{eq:eq35} 
\end{align} 
where $g$ is the acceleration due to gravity, $D$ is the molecular diffusivity and $\mu$ is the fluid dynamic viscosity.
The ratio of kinematic viscosity to solute diffusivity $D$ (or molecular diffusivity) determines the Schmidt number
\begin{equation}
\sch=\frac{\mu}{\rho_r D}\text{ ,}    
\label{eq:sc}
\end{equation}
with $\rho_r$ the average (or reference) fluid density within the domain.
Similarly, for thermally-driven flows one can define the Prandtl number ($\pr$), in which the molecular diffusion is replaced by its thermal counterpart.

Modelling heat or mass transport at the pore-scale requires to resolve the flow within the interstitial space.
Momentum transport is controlled by continuity and Navier-Stokes equations, respectively:
\begin{equation}
\nabla\cdot\mathbf{\tilde{u}}=0\text{  ,}
\label{eq:cont1}
\end{equation}
\begin{equation}
\tilde{\rho}\left[\frac{\partial \mathbf{\tilde{u}}}{\partial t}+\left(\mathbf{\tilde{u}}\cdot\nabla\right)\mathbf{\tilde{u}}\right]=-\nabla \tilde{p} +\mu\nabla^2\mathbf{\tilde{u}} + \tilde{\rho}\mathbf{g}\text{  ,}
\label{eq:ns1}
\end{equation}
where $\mathbf{\tilde{u}}$, $\mathbf{\tilde{\rho}}$ and $\tilde{p}$ are the velocity, density and pressure fields, respectively, and $\mathbf{g}$ indicates acceleration due to gravity.
We assumed that the Boussinesq approximation applies, which is reasonable for geophysical processes such as carbon sequestration~\citep{landman2007heat} (additional details on this assumption are provided in Sec.~\ref{sec:stateart}).
The fluid density $\tilde{\rho}$ is typically defined by an equation of state (EOS) that depends on both solute concentration and fluid temperature (other scalars present in the system may be similarly treated).
When linearized, the EOS may be rewritten to obtain the density $\tilde{\rho}$ as a function of temperature ($\tilde{T}$) and concentration ($\tilde{C}$) (possible limitations of this approach are discussed in Sec.~\ref{sec:stateart}).
With respect to the value $\tilde{\rho}_r$ defined at the reference state ($\tilde{C}_r,\tilde{T}_r$), it reads
\begin{equation}
\tilde{\rho}=\tilde{\rho}_r+\alpha_c(\tilde{C}-\tilde{C}_r)+\alpha_t(\tilde{T}-\tilde{T}_r),
\label{eq:eos}
\end{equation}
where $\alpha_c,\alpha_t$ are the expansion coefficients relating the density to the variations of concentration and temperature, respectively, being typically $\alpha_c>0$ and $\alpha_t<0$.
In this case, assuming the presence of solute scalars only, Eq.~\eqref{eq:eos} reduces to the form $\tilde{\rho}=\tilde{\rho}(\tilde{C})=\tilde{\rho}_r+\alpha_c(\tilde{C}-\tilde{C}_r)$.

Solute conservation is accounted by the advection-diffusion equation:
\begin{equation}
\frac{\partial \tilde{C}}{\partial t}+\mathbf{\tilde{u}}\cdot\nabla \tilde{C}=D\nabla^2 \tilde{C}.
\label{eq:ade1}
\end{equation}
Eq.~\eqref{eq:cont1}-\eqref{eq:ade1} are solved for the fluid domain to determine the evolution of the flow at the pore-scale \cite{depaoli2023convective}.
When heat transport is considered, a diffusive heat flux in the solid matrix may be also accounted \cite{lohsesun2020,gasow2020} (additional details will be provided in Sec.~\ref{sec:shpore}).
The presence of additional phases is not discussed in this review, and we refer to \cite{blunt2017multiphase} for pore-scale modelling approaches of multiphase flows.

Notwithstanding recent significant improvements of numerical schemes and computational infrastructures, resolving real-scale convective flows from the pore level to the reservoir scale requires a computational effort that is beyond the present capabilities.
To overcome this issue, a possible approach consists of modelling the flow at an intermediate scale between pores length scale and domain height, i.e., the Darcy scale.
Despite missing a precise description of the flow dynamics at the pore level, Darcy models have proved to be a reliable framework to determine the overall long-time behaviour for the transport of species in convective porous media flows \cite{huppert2014fluid,moure2023thermodynamic,lasser2023salt,anderson2022mushy}.
In the following, we will describe under which assumptions a convective flow in porous media can be modelled as a continuum via the Darcy flow approximation.

\subsection{Darcy model and dispersion}\label{sec:convflow}
\begin{figure}[t!]
    \centering
    \includegraphics[width=0.35\textheight]{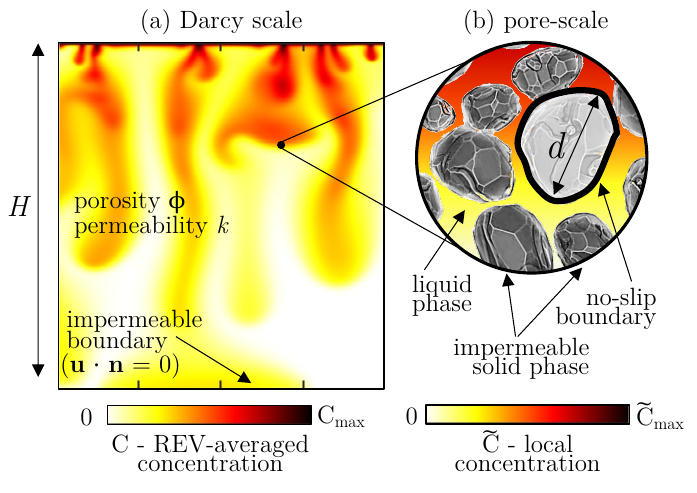}
    \caption{
Model of the flow at different scales.
(a)~At the Darcy level, all flow quantities are obtained as averaged over the REV.
Solid boundaries (in this example at the bottom of the domain) are impermeable to fluid, i.e. the velocity component perpendicular to the wall is zero ($\mathbf{n}$ is the unit vector normal to the wall).
However, slip along this boundary is possible. 
(b)~At the pore-level, the fluid phase flows within the interstitial space of the solid matrix, which is made of impermeable solid objects. 
Over the surface of each of these objects, no-slip boundary condition applies.
    }
    \label{fig:rev}
\end{figure}

A possible strategy to model flows in porous media consists of taking the average of relevant quantities (velocity, concentration and pressure fields) over a representative volume that contains several pores \citep{nield17}. 
An illustrative example is sketched in Fig.~\ref{fig:rev}.
The size of the volume (indicated as representative elementary volume, REV) over which the average is computed is larger compared to the pores length scale $d$, but still smaller than the domain reference length $H$.
The Darcy model is based on empirical observations initially proposed more than 150 years ago \cite{darcy1856fontaines}, and the later derived analytically by \citep{whitaker1986flow}.
We refer to \citep{bear2018modeling,nield17} and references therein for additional details.
The key assumption of the Darcy equation is that the average flow velocity over the representative volume is proportional to the pressure gradient applied to the volume via the fluid viscosity and a property of the medium defined as permeability.
These conditions are achieved when the flow inertia is negligible compared to viscous forces \cite{nield17}.
In the following, we will characterise the medium properties and the governing flow parameters, and finally we will discuss a model for the Darcy flow.

The characteristic geometrical properties of the solid matrix and its intimate interaction with the interstitial fluid determine the flow behaviour.
The main macroscopic parameters used to characterise a porous medium are: (i)~porosity $\phi$, defined as the ratio of volume of fluid to the total volume (fluid and solid), and (ii)~permeability $k$, a measure of the resistance opposed by the medium to the flow.
For a given porous medium, the Darcy number represents the relative importance of permeability and its cross-sectional reference area. 
With respect to the domain reference length scale $H$, the Darcy number reads:
\begin{equation}
\da=\frac{k}{H^2}\text{  .}
\label{eq:meth01} 
\end{equation}

A convective flow is driven by density differences, and therefore a possible velocity scale is the buoyancy velocity $U$, i.e. the free fall velocity of a parcel of immiscible fluid surrounded by fluid having a density contrast $\Delta\rho$, which is defined as
\begin{equation}
U=\frac{g\Delta\rho k}{\mu}\text{  .}
\label{eq:eq33} 
\end{equation} 
We observe that $U$ is independent of the any length scale, and it relates to the fluid ($\Delta\rho,\mu$), medium ($k$) and domain ($g$) properties.
In addition to the domain length scale ($H$), one can consider as a reference length scale the distance $\ell$ over which advection and diffusion balance \citep{slim2014solutal}
\begin{equation}
\ell=\frac{\phi D}{U}\text{  }
\label{eq:eq34} 
\end{equation} 
(in thermal convection, $\ell=D/U$ with $D$ representing the thermal diffusivity).
The evolution of the fluid layer is controlled by buoyancy, which tends to drive the flow towards a stable configuration, and diffusion, acting to reduce local concentration gradients and increasing the mixing of solute in the domain. 
The relative importance of the strength of these contributions is evaluated by the Rayleigh-Darcy number $\raM$
\begin{align}
\raM=\frac{H}{\ell}\text{,}
\label{eq:eq36} 
\end{align} 
obtained combining the Rayleigh number $\raT$ \eqref{eq:eq35} and the Darcy number $\da$ \eqref{eq:meth01}.
In particular, $\raM=\raT\da/\phi$ in the instance of solutal convection, with the solid being impermeable to the solute fluxes, and $\raM=\raT\da$ for thermally-driven cases in conductive media.
While in the thermal case an equilibrium between the solid and the fluid phases may be achieved, in solutal convection the solid phase is always solute-free.
Notwithstanding this difference, when thermal equilibrium locally occurs between the solid and the fluid phases, results for thermal convection can be equally interpreted as results for solutal convection, and vice-versa, provided that the Rayleigh-Darcy number is matched \cite{hewitt2020vigorous}.
The Rayleigh-Darcy number includes all the macroscopic properties of the system: domain ($g,H$), medium ($k,\phi$) and fluid ($D,\mu,\Delta\rho$) properties.
In addition, when the spatial coordinates are made dimensionless with respect to $\ell$ \eqref{eq:eq34}, $\raM$ can be interpreted as the dimensionless domain height \cite{slim2014solutal}.

A Darcy-type flow occurs when the size of the flow structures is much greater than the reference length of the REV \cite{hewitt2020vigorous}.
The reference length scale is in this case the pore-scale, which is proportional to $\sqrt{k}$.
In quantitative terms, the criterion above is fulfilled when: (i)~the pore-scale Reynolds number is small, i.e., viscous dissipation $(\mu U)$ dominates over inertia $(\rho_{r}U^{2}\sqrt{k})$, and (ii)~the smallest length scale of the flow ($\ell$) is large compared to the pore size ($\sqrt{k}$).
These conditions are matched if:
\begin{align}
\frac{\rho_{r}U^{2}\sqrt{k}}{\mu U}\ll1\Rightarrow\Rey&=\frac{\rho_{r}U\sqrt{k}}{\mu}=\frac{\raM\da^{1/2}}{Sc}\ll 1 \label{eq:eqpere} \\
\frac{\ell}{\sqrt{k}}\gg1\Rightarrow\pe&=\frac{\sqrt{k}}{\ell}=\raM\da^{1/2}\ll 1,
\label{eq:eqpere2} 
\end{align} 
i.e. when Reynolds ($\Rey$) and P\'eclet ($\pe$) numbers are much less than unity.
Note that in these definitions the pore length scale ($\sqrt{k}$) and the buoyancy velocity ($U$) are used as length and velocity scales, respectively.

We consider a fluid-saturated homogeneous and isotropic porous medium with porosity $\phi$ and permeability $k$ (Fig.~\ref{fig:rev}a) fulfilling the conditions \eqref{eq:eqpere}-\eqref{eq:eqpere2}.
The flow field is fully described by the continuity and Darcy equations, respectively:
\begin{equation}
\nabla\cdot\mathbf{u}=0\quad
\label{eq:cont2}
\end{equation}
\begin{equation}
\mathbf{u}=-\frac{k}{\mu}\left(\nabla p+\rho  \mathbf{g}\right) \text{  .} 
\label{eq:eqadim2}
\end{equation}
Note that in this case $\mathbf{u}$ is the seepage or Darcy velocity, and it represents the value of fluid velocity averaged over the REV (Fig.~\ref{fig:rev}b).
It is related to the fluid velocity averaged over the fluid phase of the REV ($\mathbf{\tilde{u}}$) via the Dupuit-Forchheimer relationship $\mathbf{u}=\phi\mathbf{\tilde{u}}$ \cite{nield17}.
Same applies to pressure $p$ and density $\rho$.

The evolution of the concentration field is controlled by the advection-diffusion equation 
\begin{equation}
\phi\frac{\partial C}{\partial t}+\nabla \cdot(\mathbf{u}C-\phi D \nabla C)=0 \text{  ,} 
\label{eq:eqadim3}
\end{equation}
where $t$ is time, $C$ is the concentration averaged over the REV and $D$ is the solute diffusivity, which is assumed constant and independent from the flow. 
In a more general formulation discussed in Sec.~\ref{sec:disp} this coefficient may be replaced by a dispersion tensor $\mathbf{D}$, that depends on the local flow conditions ($\mathbf{u}$) or the fluid properties ($Sc$).
While the solid is commonly impermeable to solute, in the thermal case a diffusive heat flux may occur within the solid matrix.
In case of thermal equilibrium between the solid and the liquid phases, Eqs.~\eqref{eq:cont2}-\eqref{eq:eqadim3} keep being valid.

\subsubsection{Dispersion}\label{sec:disp}
Solute redistribution induced by the fluid carrying the solute and flowing through the porous medium is defined as dispersion \citep{woods2015flow}. 
This mechanism, which has the effect of homogenizing the solute concentration field, adds to the contribution of molecular diffusion.
For this reason, these two contributions that are originated by very different physical mechanisms are often grouped within a unique formulation. 
In porous media, dispersion may arise due to several reasons: pore-scale change of flow direction (mechanical dispersion), heterogeneous permeability fields (large-scale dispersion) or other mechanisms, such as no-slip at the boundary of the pores or dead-end pores (anomalous dispersion).
These effects are the result of the pore-scale dynamics, and can be considerably more effective (up to few orders of magnitude) than the solute spreading due to molecular diffusion.
Therefore, it may be necessary to account for the presence of dispersion when modelling the flow at the Darcy scale. 
Here we consider the contribution of mechanical dispersion and molecular diffusion, usually grouped in a term defined as hydrodynamics dispersion.
For simplicity, in the following we will indicate this mechanism as dispersion, ad we refer to \cite{dentz2023mixing} for a general theoretical discussion on dispersion-induced mixing.

A classical approach to account for the effects of dispersion consists of replacing the molecular diffusion coefficient, $D$ in Eq.~\eqref{eq:eqadim3}, with a dispersion tensor $\mathbf{D}$ which depends on the local flow conditions.
Typically, the dispersion tensor is anisotropic and aligned with the flow, meaning that it can be decomposed into two components in the directions parallel ($D_{L}$, longitudinal dispersion) and perpendicular ($D_{T}$, transverse dispersion) to the local flow velocity $\mathbf{u}$.
This model is labelled as Fickian dispersion \cite{bear1961tensor}.
With these assumptions, the dispersion tensor takes the form:
\begin{equation}
\mathbf{D} = D\mathbf{I}+(\alpha_{L}-\alpha_{T})\frac{\mathbf{u}\mathbf{u}}{\vert\mathbf{u}\rvert}+\alpha_{T}\mathbf{u}\mathbf{I},
    \label{eq:disp_r0}
\end{equation}
where $\mathbf{I}$ is the identity tensor, and the coefficients $\alpha_{L}=D_{L}/U$ and $\alpha_{T}=D_{T}/U$ correspond to the dispersivities of the medium in longitudinal and transverse directions, respectively.
For solute transport and $\pe\gg1$, dispersion in the cross-flow direction is typically 1 order of magnitude smaller than in the stream flow direction (possible limitation of this assumption are discussed in \cite{bijeljic2007pore}).
The ratio of these two contributions is quantified by the dispersivity ratio
\begin{equation}
r=\frac{D_L}{D_T}.
    \label{eq:disp_r}
\end{equation}
The magnitude of $D_L$ and $D_T$ is estimated with the aid of correlations based on experiments and simulations \cite[see][ and references therein]{woods2015flow}.
Longitudinal and transverse dispersion coefficients depend on many parameters, namely Schmidt number \cite{delgado2001measurement}, Reynolds number \cite{wood2007inertial}, tortuosity of the medium \citep{delgado2007longitudinal}, P\'eclet number \citep{alkindi2011investigation}, fluid phases \citep{delgado2007longitudinal}. 
We refer to \cite{eskandari2019investigation} for a review of numerical, experimental and theoretical works in this field.

\begin{figure}
    \centering
    \includegraphics[width=0.35\textheight]{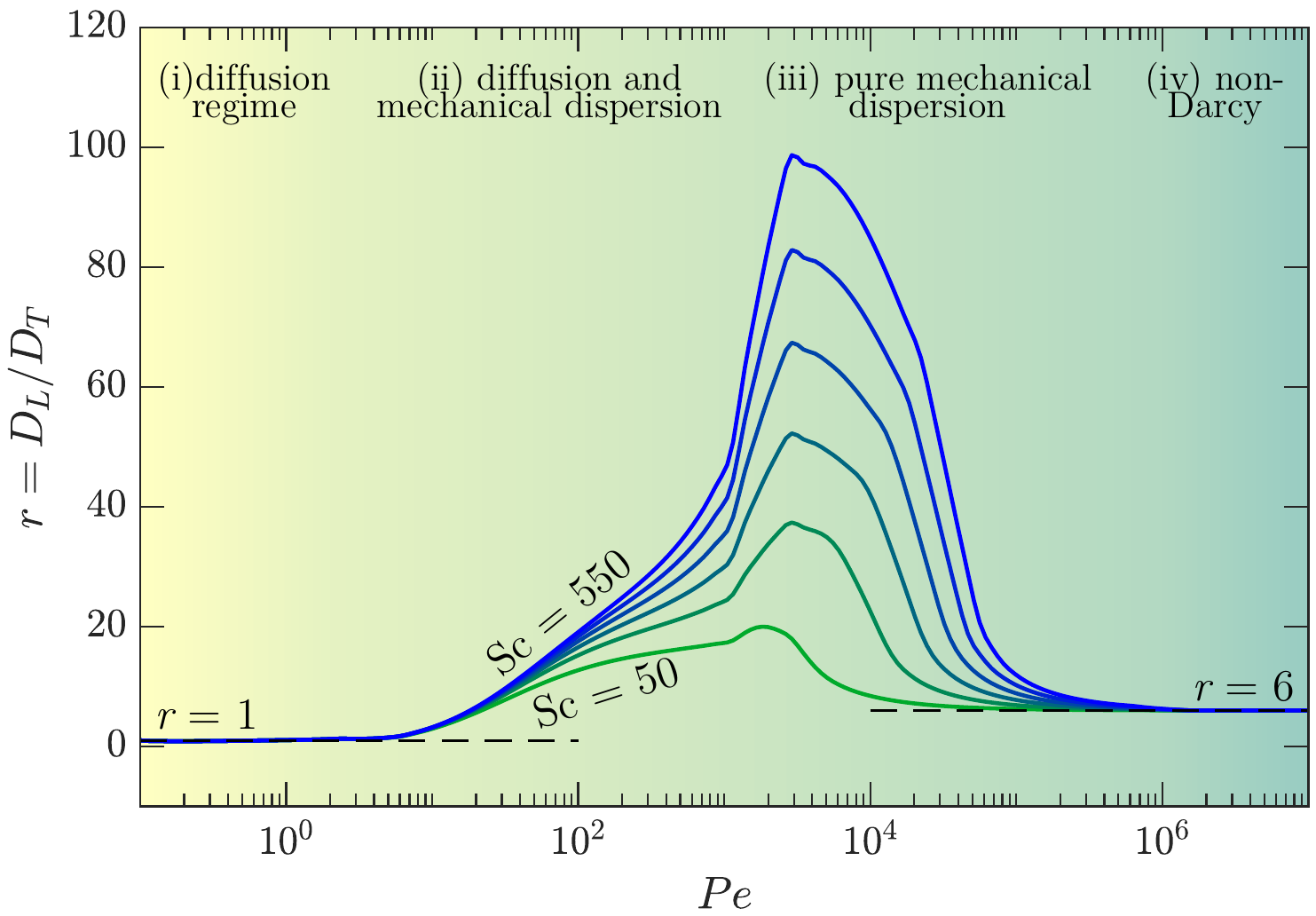}
    \caption{
    Dispersivity ratio $r=D_L/D_T$ shown for porosity $\phi=0.37$, tortuosity $\tau=0.68$ and different Schmidt numbers, namely $\sch = 50$, 150, 250, 350, 450 and 550.
    The correlations proposed by \citep{delgado2007longitudinal} have been employed.
    The advective flow is divided in several regimes, discussed in the text.
    }
    \label{fig:dispersion}
\end{figure}

We consider here an example of a medium composed of monodispersed beads, typical of numerical and experimental setups commonly employed in pore-scale investigations, and we show that the dispersivity ratio $r$ may considerably vary as a function of the P\'eclet number of the flow
(a similar procedure applies for different media -- porosity, tortuosity -- and flow -- P\'eclet number -- properties).
We consider the case of solutal convection in a monodispersed bead pack at $\sch\le550$.
For a monodispersed close random packing the porosity is $\phi = 0.37$ \citep{dullien2012porous,haughey1969structural} and the tortuosity (the ratio of actual flow path length to the straight distance between the ends of the flow path \cite{bear2018modeling}) is $\tau=0.68$ \cite[see][ and references therein]{vafai2005handbook}.
We use the empirical correlations proposed by \citep{delgado2007longitudinal}, obtained for laboratory experiments, i.e., in architecture-controlled media, whereas we refer to \cite{gelhar1992critical} for a review of dispersion relations in field-scale data.
The results proposed by \citep{delgado2007longitudinal} are valid for liquids and at $\sch\le550$, and we report the dispersivity ratio in Fig.~\ref{fig:dispersion} (for $\sch>550$, similar correlations are provided).
Four main flow regimes have been identified, for increasing $Pe$:
(i)~diffusion regime, with molecular diffusion being the dominant mechanism; 
(ii)~diffusion and mechanical dispersion, when the two contributions are comparable;
(iii)~pure mechanical dispersion, when the influence of molecular diffusion is negligible;
(iv)~non-Darcy, when the effects of inertia and turbulence cannot be neglected.
Note that these correlations are obtained from experimental measurements performed for a wide parameters space, and a sharp separation between these regimes is hard to identify.
A theoretical prediction is available for low ($r=1$, \cite{koplik1988transport}) and high ($r=6$, \cite[][ and references therin]{delgado2007longitudinal}) P\'eclet numbers.
A similar regime classification has been also proposed by \cite{perkins1963review}.

With this example we have shown that in general $r$ varies with $Pe$ and $Sc$ among the other parameters, and also across the scales \cite{majdalani2015solute}. 
To simplify the picture, a possible approach used in numerical simulations consists of fixing the values of $D_{L}$ and $D_{T}$ or $r$ \citep{liang2018effect}, which is a reasonable approximation if a narrow range of $Pe$ is considered.
Results on the effect of dispersion on convective flow are presented in Sec.~\ref{sec:dispgran}.

\subsection{Flow configurations and quantification of mixing}\label{sec:convconv}
Convective processes of practical interest are characterised by the mixing of one or more scalar quantities (e.g., the concentration of a dispersed solute phase) in the ambient fluid, and predicting the time required to achieve a certain degree of mixing may be necessary.
In the instance of geological carbon sequestration, for example, it is desired to find the time required to dissolve a considerable fraction of the CO$_{2}$ injected, to assess the reliability of a given sequestration site. 
These estimates can be obtained via experiments and simulations in representative flow configurations, which are well controlled and designed to reproduce the main features observed in environmental and industrial cases.
In this Section, we will first introduce the main flow configurations investigated in literature, with clear indication of the initial and boundary conditions.
Then we will define a general framework and identify relevant observables required to quantify the mixing and analyze the evolution of the system.

\begin{figure}[h!]
\center
\includegraphics[height=0.70\textheight]{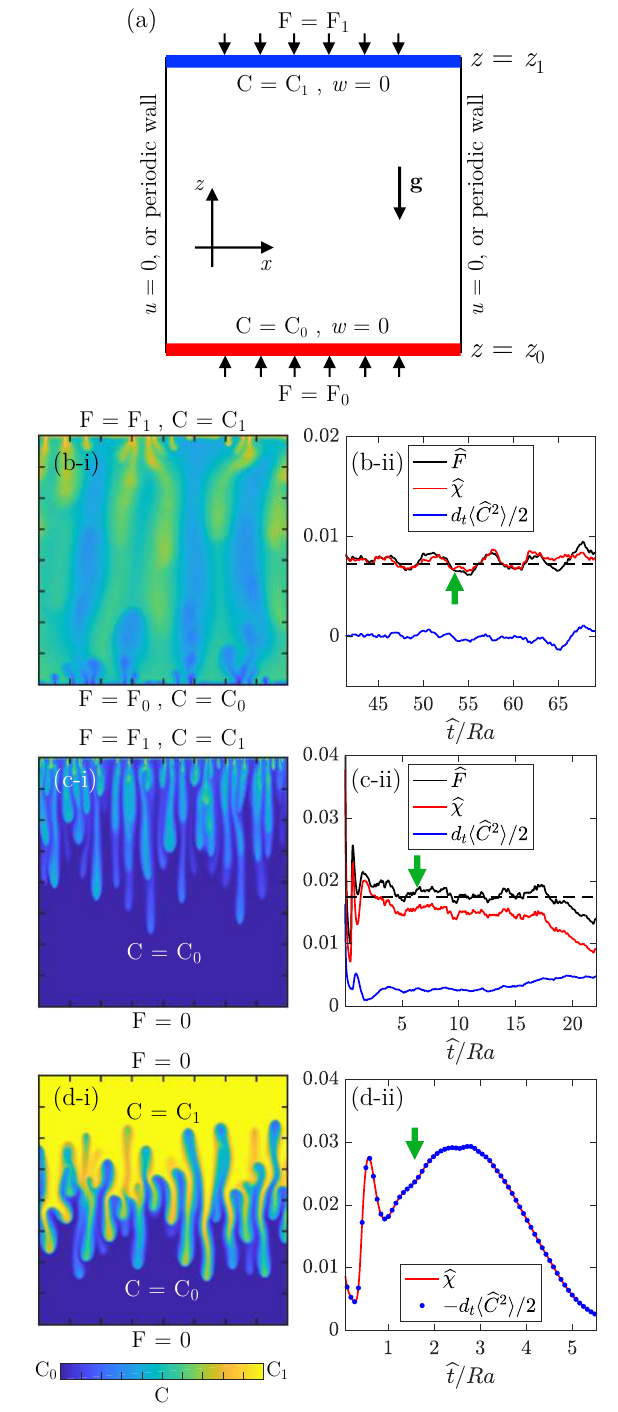}
\caption{\label{fig:comparsion}
Flow configurations 
\cite[adapted with permission from][]{depaoli2019universal}.
(a)~Sketch of boundary conditions applied at the top (label 1) and bottom (label 0) boundaries in terms of flux $F$ and concentration $C$.
All boundaries are impermeable to fluid ($\mathbf{u}\cdot\mathbf{n}=0$), and side boundaries may be also considered as periodic.
The reference frame ($x,z$) and gravity ($\mathbf{g}$) are also indicated. 
Three flow configurations are shown: (b) Rayleigh-B\'enard, (c) one-sided and (d) Rayleigh-Taylor. 
An exemplar field obtained for two-dimensional simulations at $\raM=7244$ is reported.
The field is taken at the time indicated by the green arrows in panels (b-ii), (c-ii) and (d-ii), where the evolution of the parameters $\widehat{\chi}$, $\widehat{F}$ and $d_{t}\langle C^{2}\rangle/2$ is reported (the operator $d_{t}$ stands for the time-derivative).
Quantities are computed as in Eq.~\eqref{eq:global5} and made dimensionless with respect to the length-scale $\mathcal{L}=\ell$.
The time-averaged value of $\widehat{F}$ is also shown (dashed lines) in panels~(b-ii) and (c-ii).
}
\end{figure}

Three archetypal flow configurations are generally employed to investigate the dynamics of convection in porous media.
They consist of analogue systems that help us to have a comprehension of specific scenarios occurring in nature.
A sketch to illustrate possible boundary conditions applied is shown in Fig.~\ref{fig:comparsion}(a). 
At the top (label 1) and bottom (label 0) boundaries, both flux $F$ and concentration $C$ may be prescribed (the flux will be defined more precisely later in this section).
All boundaries are considered impermeable to fluid, i.e. no penetration condition applies ($\mathbf{u}\cdot\mathbf{n}=0$, being $\mathbf{u}$ the fluid velocity and $\mathbf{n}$ the vector perpendicular to the boundary. 
However, periodic conditions on the side boundaries may be also considered for convenience in numerical studies, with no difference on the modelling described in the following.
In all cases considered here, the domain boundaries are assumed impermeable to fluid, and the fluid is supposed initially still [$\mathbf{u}(t=0)=0$].
The boundary conditions for the solute (fixed concentration or no flux) will determine the nature of the system considered (steady or transient), whereas the initial condition for the solute (uniform concentration, or two fluid layers with different concentration) will control the flow evolution.
The flow configurations considered are:
\renewcommand{\theenumi}{\roman{enumi}}
\begin{enumerate}
\item Rayleigh-B\'enard (Fig.~\ref{fig:comparsion}b-i): the solute concentration is fixed at the horizontal boundaries, so that the density of the fluid at the bottom wall ($C=C_{0}$) is lighter than the density of the fluid at the top wall ($C=C_{1}$) \cite{otero2004high,hewitt2012ultimate}. 
This unstable flow attains a statistically steady state, which is rigorously steady for sufficiently low Rayleigh-Darcy numbers \cite{graham1994plume}. 
A scalar flux is possible through the upper ($F=F_{1}$) and the lower ($F=F_{0}$) boundaries.
\item One-sided (Fig.~\ref{fig:comparsion}c-i): the concentration is imposed at the upper wall, where a solute flux is also possible ($C=C_{1}, F=F_{1}$), and the domain is impermeable to solute at the lower wall ($\partial C/\partial z=0$, corresponding to $F_{0}=0$). 
This configuration originates a time-dependent flow, and the domain, initially filled with uniform solute concentration $C=C_0$, is gradually filled with the solute coming from the upper boundary \cite{hesse2012phd,pau2010high,slim2014solutal}. 
\item Rayleigh-Taylor (Fig.~\ref{fig:comparsion}d-i): both walls are impermeable to the scalar ($F_{0}=F_{1}=0$). 
The domain initially consists of two uniform layers of different density ($C=C_{1}$ for the upper portion, and $C=C_{0}$ for the lower portion), so that the flow configuration is unstable \cite{depaoli2019prf,depaoli2022strong}.
Solute mixing evolves controlled by the dynamics of the flow structures.
\end{enumerate}

The flow configurations considered differ in terms of boundary conditions and evolution, and suitable flow observables are required to estimate the mixing state of each system. 
For instance, the Sherwood number $\sh$, defined as the ratio of the convective to the diffusive mass transport, is suitable in solute-permeable domains (e.g., the Rayleigh-B\'enard case), but it does not provide any indication in closed domains (e.g., the Rayleigh-Taylor case).
Therefore, in each flow configuration different quantities are used, which are related through exact mathematical relations that are derived here. 
Following \citep{hidalgo2012scaling}, we take the advection-diffusion equation~\eqref{eq:eqadim3} multiplied by $C$, and we integrate over the entire domain. 
We use the hypothesis of incompressibility of the flow \eqref{eq:eqadim2} together with the impermeability of the boundaries to the fluid (note that the same result is achieved assuming periodicity in horizontal direction).
After some algebraic manipulations, we obtain the following exact global relation:
\begin{equation}
\frac{\phi}{2}\frac{d \langle C^{2}\rangle}{d t} = \frac{\phi}{H}\left( C_{1}F_{1}+C_{0}F_{0}\right)-\phi \chi\text{  ,}
\label{eq:global1}
\end{equation}
where $\langle\cdot\rangle$ indicates the volume average.
Eq.~\eqref{eq:global1} relates the mean squared concentration, the solute flux through the walls $F$ and the mean scalar dissipation within the domain $\chi$, respectively defined as
\begin{equation}
    F_i=\frac{D}{L}\int_{0}^{L}\frac{\partial C}{\partial z}\biggr\rvert_{z=z_i}\text{ d}x\quad\text{with $i=\{0,1\}$}    \text{  ,}
    \label{eq:fff3}
\end{equation} 
with $L$ domain width, and
\begin{equation}
    \chi = D\langle\lvert\nabla C\rvert^{2}\rangle \text{  .}
    \label{eq:diss33}
\end{equation} 
When $C$ is defined as a mass concentration, $F$ may be interpreted as the average mass of solute that enters (or leaves) the domain per unit of surface area and time.
Eq.~\eqref{eq:global1} can be interpreted as follows.
The rate of change of mean squared concentration within the domain is the result of external contributions ($F_{0},F_{1}$, either positive or negative) and dissipation of mixing energy ($\chi$ is always positive, therefore it contributes to a reduction of scalar variance $\langle C^{2}\rangle$).

To enable comparisons among different systems, a possible set of dimensionless variables consists of $\mathcal{L}$ for lengths, $\phi \mathcal{L}/U$ for time, and $U$ for velocities.
The concentration $C$ is made dimensionless as $\widehat{C}=(C-C_0)/\Delta C$,
where $\widehat{\cdot}$ indicates dimensionless quantities and $\Delta C = C_1-C_0$.
Making Eq.~\eqref{eq:eqadim3} dimensionless with these variables and proceeding as above, we obtain a dimensionless form of Eq.~\eqref{eq:global1} that reads:
\begin{equation}
\frac{1}{2}\frac{d \langle \widehat{C}^{2}\rangle}{d \widehat{t}} = \frac{\phi D}{U\mathcal{L}}\left( \widehat{F}-\widehat{\chi}\right)
\label{eq:global2}
\end{equation}
with $\widehat{F}=F_1\mathcal{L}/(D\Delta C)$ the dimensionless flux and $\widehat{\chi}=\chi \mathcal{L}^2/[D(\Delta C)^{2}]$ the dimensionless mean scalar dissipation.
Note that in this expression the contribution of the flux at the bottom boundary vanishes, due to the set of dimensionless variables considered.
The reference length scale $\mathcal{L}$ has not been defined yet and it can be conveniently set in each configuration.
With respect to the systems previously introduced, the following scenarios appear:
\begin{enumerate}
\item Rayleigh-B\'enard (Fig.~\ref{fig:comparsion}b-ii): after an initial transient phase, the system attains a statistically steady state \cite{hewitt2012ultimate,pirozzoli2021towards}.
The time-average of Eq.~\eqref{eq:global2} returns
\begin{equation}
\overline{\widehat{F}} = 
\overline{\widehat{\chi}},
\label{eq:global3}
\end{equation}
where $\overline{\cdot}$ indicates the time-averaging operator.
We observe in Fig.~\ref{fig:comparsion}(b-ii) that while a non-zero instantaneous contribution $d \langle \widehat{C}^{2}\rangle/d \widehat{t}$ is present, $\overline{\widehat{F}}$ and $\overline{\chi}$ fluctuate around their time-averaged value (black dashed line).
Note that the reference length scale normally used in this configuration is $\mathcal{L}=H$, which gives in \eqref{eq:global2} the prefactor $\phi D/(U\mathcal{L})=1/\raM$.
The quantity used to evaluate the mass transfer in this configurations is the Sherwood number
\begin{equation}
\sh=\frac{H}{\Delta C L}\overline{\int_{0}^{L}\frac{\partial C}{\partial z}\biggr\rvert_{z=z_1}\text{ d}x}\text{  ,} 
    \label{eq:shw1}
\end{equation}
defined as the relative contribution of convective and diffusive to diffusive mass transport.
Using the definition of $\widehat{F}$ and Eq.~\eqref{eq:global3}, $\sh$ can be related to the flux and the mean scalar dissipation \cite{otero2004high}:
\begin{equation}
\sh=\raM\overline{\widehat{F}}=\raM\overline{\chi}.
\label{eq:global4}
\end{equation}

\item One-sided (Fig.~\ref{fig:comparsion}c-ii): the domain is impermeable to solute at the lower wall ($F_{0}=0$).
By setting $\mathcal{L}=\ell$ as defined in \eqref{eq:eq34}, Eq.~\eqref{eq:global2} is independent of $\raM$ and reads
\begin{equation}
\frac{1}{2}\frac{d \langle \widehat{C}^{2}\rangle}{d \widehat{t}} = \widehat{F}-\widehat{\chi},
\label{eq:global5}
\end{equation}
where 
\begin{equation}
\widehat{F}=\frac{\phi D}{U\Delta C}\frac{1}{L}\int_{0}^{L}\frac{\partial C}{\partial z}\biggr\rvert_{z=z_1}\text{ d}x .
\label{eq:global5b}
\end{equation}
This choice for $\mathcal{L}$ is convenient to compare systems having different $\raM$ because the value of $\widehat{F}$ appears to be universal, as will be later discussed in Sec.~\ref{sec:os}.
The time-dependent flow originated from this configuration consists of three-main flow regimes \cite{slim2014solutal,depaoli2017solute,croccolo2022}. 
Initially ($\widehat{t}<10^{3}$) diffusion dominates and a high-concentration high-density unstable fluid layer thickens.
At a later stage ($\widehat{t}<16\raM$), convection takes place and plumes formed at the top boundary layer grow and invade the domain. 
In this phase $\widehat{F}$ is statistically steady and characterized by a value (black dashed line) that is independent of the Rayleigh-Darcy number considered.
A similar behaviour holds for $\chi$, but a closer inspection reveals that after the fingers reach the bottom ($\widehat{t}>10\raM$) an increase of $d \langle \widehat{C}^{2}\rangle/d \widehat{t}$ is observed.
A corresponding decreasing behaviour is reflected in $\chi$, but with half the amplitude.
After the upper layer of the domain is also saturated ($\widehat{t}>16\raM$), the dissolution flux $\widehat{F}$ drops, and the system enters the shutdown regime. 
\item Rayleigh-Taylor (Fig.~\ref{fig:comparsion}d-ii): the domain is impermeable to the solute ($\widehat{F}=0$) and Eq.~\eqref{eq:global2} reads
\begin{equation}
\frac{1}{2}\frac{d \langle \widehat{C}^{2}\rangle}{d \widehat{t}} =-\frac{\phi D}{U\mathcal{L}} \widehat{\chi}.
\label{eq:global6}
\end{equation}
The flow is initialised considering two fluid layers of different density in an unstable configuration. 
Eq.~\eqref{eq:global6} suggests that all the potential energy initially stored by keeping the two phases segregated is dissipated as time evolves. 
Both $\ell$ and $H$ can be considered as reference length scales, depending on which part of the flow evolution is considered.
However, \cite{depaoli2019universal} have shown that $\mathcal{L}=\ell$ provides a universal picture for the evolution of $\widehat{\chi}$, and results are presented in Fig.~\ref{fig:comparsion}(d-ii) using this length-scale.
Similarly to what observed in the one-sided configuration, the flow is initially controlled by diffusion ($\widehat{t}<10^{3}$).
Afterwards ($10^{3}<\widehat{t}<\raM/2$) the formation of fingers is observed, which merge and grow, accelerating mixing.
In this phase, occurring at $\raM<\widehat{t}<3\raM$ in the simulations considered, $\widehat{\chi}$ is observed to increase in Darcy simulations, as shown in Fig.~\ref{fig:comparsion}(d-ii), whereas it decreases in pore-scale simulations, due to finite-size effects \cite{depaoli2023convective}.
The limits in which these regimes set in are indicative, as the flow evolution in strongly influenced by the initial perturbation.
When the domain is nearly saturated, a stable density profile is achieved, local concentration gradients are not sufficient to sustain convection, which is in turn overcome by diffusion.
Correspondingly, scalar dissipation is observed to reduce, asymptotically attaining a zero value in correspondence of a uniformly mixed domain.
\end{enumerate}

A major proportion of recent studies focused on the determination of correlations of the mixing parameters ($\widehat{F}, \sh$ or $\widehat{\chi}$) with the flow parameter ($\raM$).
These results will be reviewed Secs.~\ref{sec:res2} and \ref{sec:os} for the Rayleigh-B\'enard and the one-sided configurations, respectively.

\section[Rayleigh-B\'enard convection]{Rayleigh-B\'enard convection}\label{sec:res2}
Rayleigh-B\'enard convection produces the statistically-steady flow discussed in Sec.~\ref{sec:convconv}, with mass transfer properties quantified by the Sherwood number, a time-averaged ratio of total (convective and diffusive) to diffusive mass transport at the boundaries of the domain defined in Eq.~\eqref{eq:shw1}.
Alternatively, the Nusselt number is used in case of thermal convection.
In this section we will review the results relative to Darcy and pore-scale flows in this configuration.

\subsection{Darcy flow}\label{sec:shdarcy}
\begin{figure}[t!]
\center
\includegraphics[width=0.35\textheight]{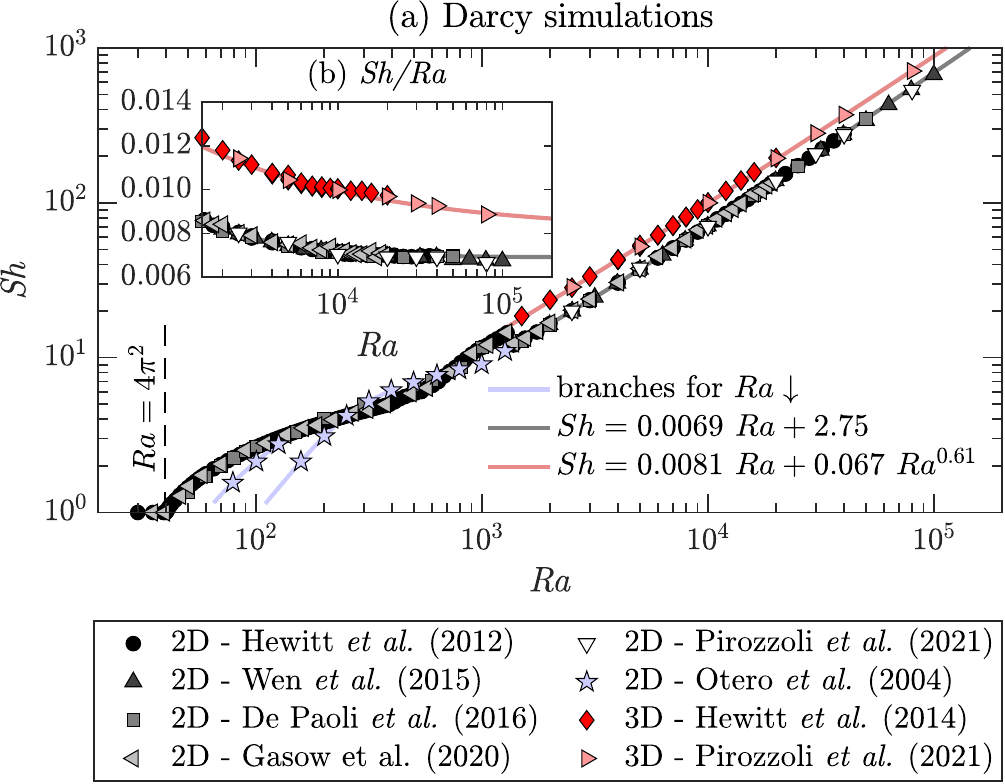}
\caption{\label{fig:sh1}
Darcy simulations.
(panel~a)~Sherwood number ($\sh$) as a function of Rayleigh-Darcy number ($\raM$), and (panel~b) in compensated form ($\sh/\raM$).
Results reported are obtained in two- \citep{hewitt2012ultimate,wen2015structure,depaoli2016influence,gasow2020} and three-dimensional \citep{hewitt2014high,pirozzoli2021towards} simulations of homogeneous and isotropic porous media.
Best fitting laws at high-Rayleigh-Darcy numbers for two- (black solid line) and three-dimensional flows (red solid line), respectively Eq.~\eqref{nu_fit2D} and Eq.~\eqref{nu_fit3D}, are also reported.
In panel (a), a subset of the two-dimensional data of \cite{otero2004high} (blue stars) is also shown to mark the presence of hysteresis effects.
The solution obtained at $\raM=1255$ was used as initial condition for these simulations.
As $\raM$ is decreased, the system evolves on two branches (blue lines), both differing from the solution obtained for increasing $\raM$.
}
\end{figure}

In the Darcy case [Eqs.~\eqref{eq:cont2}-\eqref{eq:eqadim3}], the system is uniquely controlled by the Rayleigh-Darcy number $\raM$, defined in Eq.~\eqref{eq:eq36}, which sets the flow structure.
The behaviour of $\sh$ with $\raM$ is reported in Fig.~\ref{fig:sh1} for Darcy studies available in literature for two- \cite{hewitt2012ultimate,wen2015structure,depaoli2016influence,gasow2020} and three-dimensional \citep{hewitt2014high,pirozzoli2021towards} simulations.
We briefly recall here the main features of the flow, and we refer to \cite{hewitt2020vigorous} for a detailed review of the flow structure.
For $\raM<4\pi^{2}$, the mass transport is purely diffusive \cite{horton1945convection,lapwood1948convection} and no convective motion arises ($\sh=1$).
The flow is maintained quiescent by the dissipative (diffusive) effects that dominate over convection. 
For increasing $\raM$, instabilities appear in the form of steady convective rolls \cite{graham1994plume} with corresponding increase of the convective mass transfer.
When $\raM\approx 400$, unsteady boundary layer instabilities take place and become progressively dominant.
When the driving force is sufficiently large, namely at $\raM\approx 1300$ and  $\raM\approx 1700$ for two- and three-dimensional systems \cite{otero2004high,hewitt2014high}, respectively, these instabilities turn into a dynamic formation of small plumes at the boundary layer, which eventually grow and merge into larger plumes spanning the entire domain height.
In this stage, the flow enters the \emph{high-$\raM$} regime \citep{hewitt2012ultimate}.
The dynamics described above is similar in two- and three-dimensional domains.
However, in the three-dimensional case the flow pattern obtained at low $\raM$ may be affected by the initial condition, i.e., different flow structures are obtained starting from different initial concentration distributions.
In addition, hysteresis effects have been observed in the two-dimensional case \cite{otero2004high}: when the flow is initialised using a solution obtained at higher $\raM$, the flow structure (number of rolls) and the transport properties ($\sh$) differ from those obtained starting from, e.g., a linear temperature distribution or from a solution obtained at lower $\raM$.
As a starting point, $\raM=1255$ is used by \cite{otero2004high}, and $\raM$ is progressively decreased.
The system evolves following two distinct branches (Fig.~\ref{fig:sh1}a), both differing from the solution obtained for increasing $\raM$.

Determining the scaling of $\sh$ with $\raM$ in the high-$\raM$ regime has been object of active investigation in recent years, also due to the improvements of computational capabilities.
In the frame of free fluids (i.e., no porous medium), \cite{malkus_54} and \cite{howard_64} proposed that at sufficiently high Rayleigh numbers the interior of the domain is well mixed, and the temperature gradients are localised at the wall boundary layers.
The Sherwood number is then obtained as a result of the diffusive heat flux across these layers, which is inversely proportional to their thickness, and for porous media it is predicted to scale linearly with $\raM$, $\sh \sim \raM$.
An accurate phenomenological description of the flow and scaling arguments is provided by \cite{hewitt2020vigorous}.
The linear scaling best fitting the two-dimensional numerical results \citep{hewitt2012ultimate} is
\begin{equation}
\sh = 0.0069 \raM +2.75\label{nu_fit2D}, 
\end{equation}
and it agrees also with the best known theoretical upper bound, for which $\sh\le 0.0297\raM$  \cite{otero2004high}.
The asymptotic scaling proposed by \cite{hewitt2012ultimate} [solid black line in Fig.~\ref{fig:sh1}(a)] fits well the numerical results, and it is in agreement with the above mentioned linear predictions: The compensated Sherwood number [Fig.~\ref{fig:sh1}(b)] approaches in this case the asymptotic value (0.0069).

In three-dimensional domains the situation differs, as the compensated Sherwood number has not reached yet the asymptotic linear scaling [Fig.~\ref{fig:sh1}(b)]. 
The best-fitting is in this case provided by \citep{pirozzoli2021towards}
\begin{equation}
 \sh=0.0081\raM + 0.067 \raM^{0.61},
\label{nu_fit3D}
\end{equation}
which consists of a linear relation with sublinear corrections.
The discrepancy existing between the scaling obtained in three-dimensional porous media and the linear asymptotic prediction for $\raM\to\infty$ is due to the different flow structure produced by the additional degree of freedom provided by the third spatial dimension, i.e., the flow has not reached yet the asymptotic state.
It was estimated \citep{pirozzoli2021towards} that in three-dimensional domains the asymptotic regime sets in at $\raM \approx 5 \times 10^5$, i.e. more than one order of magnitude beyond the threshold identified in two-dimensional flows, and further investigations at $\raM\approx10^{6}$ are required to confirm this finding. 

Resolving the flow equations at the Darcy scale at large $\raM$ may require extensive computational resources \cite{pirozzoli2021towards,depaoli2022strong}.
An interesting approach proposed to overcome this obstacle consists of a new modelling strategy labelled as large-mode simulation (LMS) \cite{jenny2014scale}.
With the aid of a scale-analysis, \cite{jenny2014scale} observed that: 
(i)~large-scale structures are responsible for the bulk of the production of concentration variance, 
(ii)~variance dissipation is dominated by the small diffusive scales, and
(iii)~both production and dissipation rates are independent of the Rayleigh-Darcy number.
On this ground, they propose a LMS model in which closure is achieved by replacing the actual diffusivity with an effective one, in analogy with large eddy simulations for turbulent flows. 
LMS is based on resolving the low-wavenumber dynamics only, whereas the effect of the unresolved scales on the large ones is modelled.
Results obtained with this new strategy are promising to enable simulations for long-term predictions of convective porous media flows in practical settings.

\subsection{Pore-scale flow}\label{sec:shpore}
Recent developments in computational methods allowed numerical solution of pore-resolved convective flow models, defined by Eqs.~\eqref{eq:cont1}-\eqref{eq:ade1}.
Unlike the Darcy case, in pore-scale problems the flow properties cannot be lumped into a single governing parameter, and the contribution of several flow features has to be considered.
With respect to the medium, obstacles shape and arrangement determine the medium permeability.
The medium conductivity influences heat transport through the solid phase, and the volume fraction of solid sets the porosity.
Concerning the fluid and the scalar transported, kinematic viscosity and diffusivity set the relative thickness of thermal and kinematic boundary layers [measured by the Schmidt number $\sch$, defined in \eqref{eq:sc}], while the density difference produced by the scalar determines the driving force [measured by the Rayleigh number $\raT$, defined in \eqref{eq:eq35}].
A key quantity to consider is the relative size of the pore-space to the flow structure, which determines the penetration of the buoyant plumes, responsible of convective mixing, in the domain. 
The influence of these flow parameters on the convective transport efficiency, measured by $\sh$, is discussed here.

\begin{figure}[t!]
\center
\includegraphics[width=0.35\textheight]{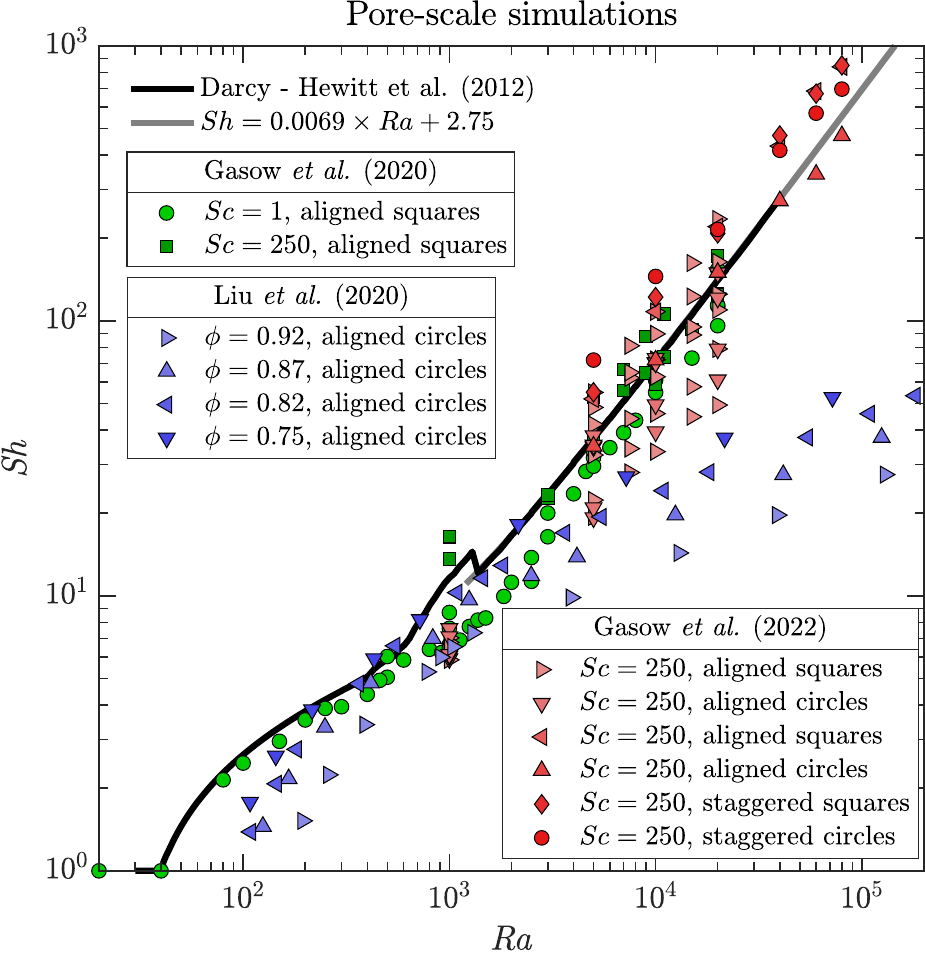}
\caption{\label{fig:sh2}
Sherwood number ($\sh$) as a function of Rayleigh-Darcy number ($\raM$) for two-dimensional pore-scale simulations.
Results refer to solutal convection \citep{gasow2020,gasow2022prediction} (i.e., solid impermeable to solute, green and red symbols) and thermal convection \citep{lohsesun2020} (i.e., conductive medium, blue symbols).
Results of two-dimensional Darcy simulations \citep{hewitt2012ultimate} (black solid line) and high-$\raM$ scaling [Eq.~\eqref{nu_fit2D}, grey solid line] are also shown.}
\end{figure}

We initially consider a solid phase impermeable to the scalar, e.g. the case of solute convection.
Accurate two-dimensional pore-scale simulations of Rayleigh-B\'enard solutal convection are presented by \cite{gasow2020}, where the porous medium is modelled as a matrix of aligned squares. 
They explored different values of porosity ($0.36\le\phi\le0.56$) and Schmidt numbers ($\sch=1$ and $\sch=250$).
The results, reported in Fig.~\ref{fig:sh2} (green symbols) as measurements of Sherwood number, indicate fair agreement with two-dimensional Darcy simulations (black solid line, \cite{hewitt2012ultimate}).
However, the pore-induced dispersion, which may be as strong as buoyancy, affects the flow structure and consequently $\sh$, and the scaling $\sh(\raM)$ appears sublinear when the porosity is increased ($\phi=0.56$).
At a low Schmidt numbers ($\sch = 1)$, pore-scale effects on the flow structure, e.g. wavenumber or width of the plumes, are qualitatively similar to those at high Schmidt numbers ($\sch = 250)$.
In a complementary study, \citep{gasow2022prediction} investigated large Schmidt numbers ($\sch = 250)$ convection, while focusing on the role of the medium properties.
Results of \citep{gasow2022prediction} are reported in Fig.~\ref{fig:sh2} (red symbols), and indicate that the dissolution coefficient depends on $\raM$ as
\begin{equation}
 \sh=1+a\raM^{1-0.2\phi^{2}} ,
\label{eq:shraphi}
\end{equation}
where $a = 0.011 \pm 0.002$ is a pore-scale geometric parameter depending on shape and arrangement of the obstacles.
The difference with respect to the Darcy case [simulations by \cite{hewitt2012ultimate} -- black line, asymptotic best fit -- Eq.~\eqref{nu_fit2D}] is apparent, as it seems that within this range of parameters, systems with same $\raM$ (achieved with different values of porosity) exhibit very different convective transport properties.

An additional degree of freedom is introduced by allowing a flux of scalar through the solid matrix, which may be the case for thermal convection.
The flow structure and the heat transfer coefficient are determined by the relative size of thermal length scale (boundary layer thickness) and porous length scale (average pore space).
These properties control the penetration of the plumes into the boundary layer region, which in turn determines the heat or mass transfer rate.
This physical mechanism has been described by three-dimensional pore-scale simulations of few pore spaces \cite{chakkingal2019numerical}. 
Later, in a complementary experimental study, \cite{ataei2019experimental} observed that while at low Rayleigh numbers the transport mechanism is less efficient than in free fluids Rayleigh-B\'enard convection, at larger Rayleigh numbers the classical scaling derived for free fluids \citep{grossmann2000scaling,grossmann2001thermal} is recovered. 
The nature of this transition has been investigated in detail by \cite{lohsesun2020} (blue symbols in Fig.~\ref{fig:sh2}).
\begin{figure}[t!]
\centering
\includegraphics[width=0.35\textheight]{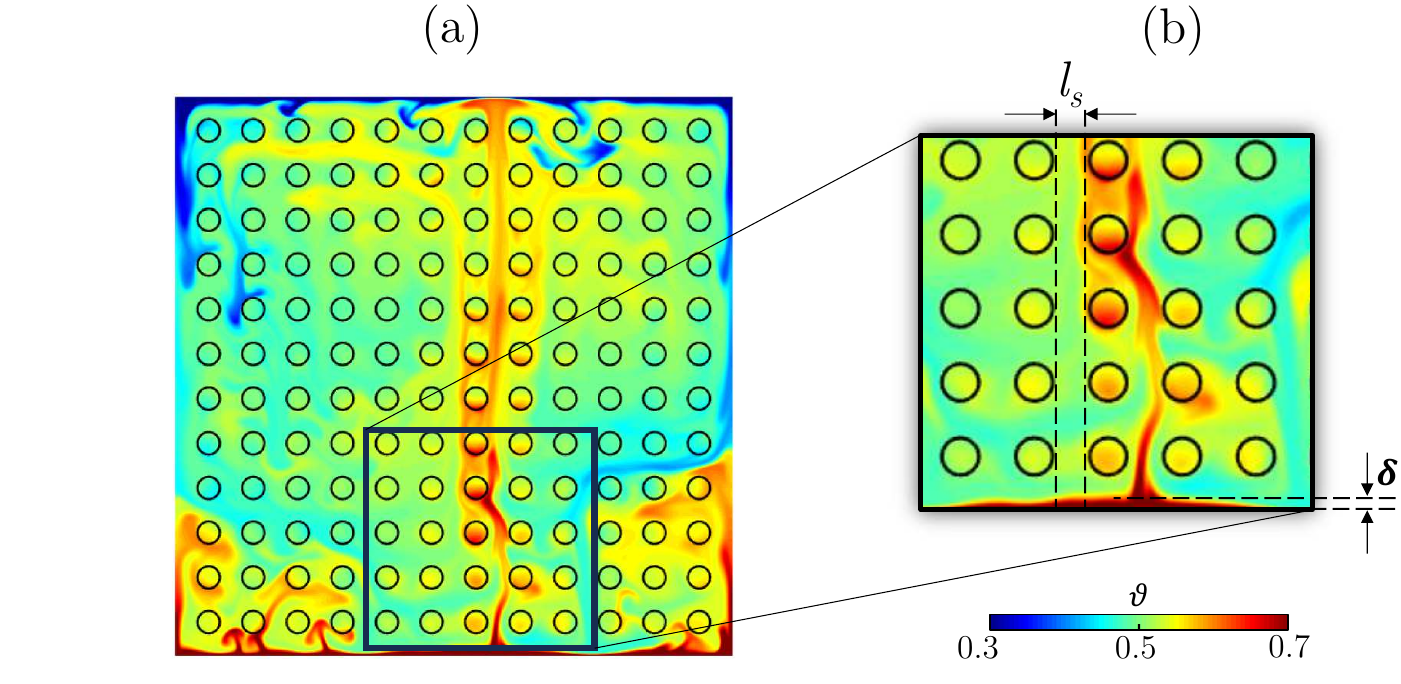}
\includegraphics[width=0.35\textheight]{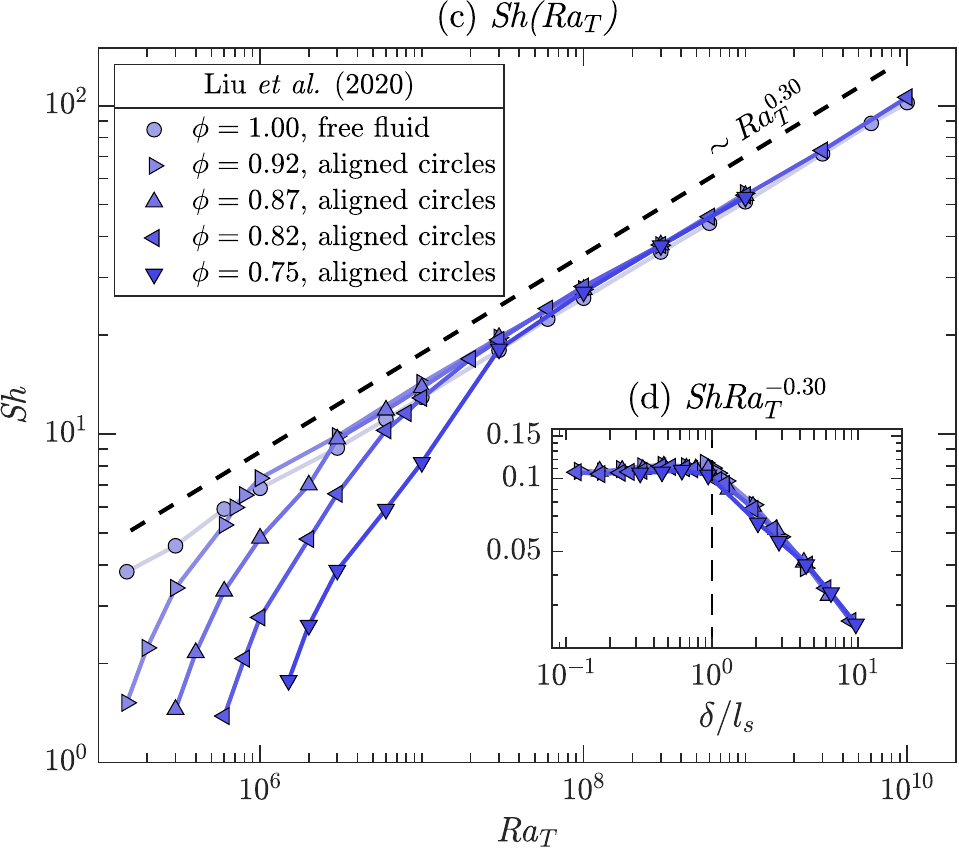}
\caption{\label{fig:sh3}
Pore-scale two-dimensional simulations \cite{lohsesun2020}.
(a)~Exemplar dimensionless temperature field ($\vartheta$)  
\cite[adapted with permission from][]{lohsesun2020}, being 0 and 1 the temperature values at the top and bottom boundaries, respectively.
(b)~Detail with explicit indication of the boundary layer thickness ($\delta=H/(2\sh)$) and the average pore scale ($l_{s}$).
The medium consist of aligned circular and conductive obstacles for Schmidt number $\sch=4.3$.
(c)~Sherwood number ($\sh$) is reported as a function of the Rayleigh number ($\raT$) for different values of porosity, $\phi$. 
Note that results for unconfined fluids ($\phi=1$) are also shown.
(d)~compensated Sherwood number ($\sh/\raT^{-0.3}$) as a function of $\delta/l_{s}$.
}
\end{figure}
Within the frame of conductive media, two-dimensional direct numerical simulations have been used to investigate the microscale flow field at $\sch=4.3$.
The obstacles consist of circles arranged in a regular manner.
When the arrangement is not regular (not shown in Fig.~\ref{fig:sh2}), a slight decrease of $\sh$ is observed.
In Fig.~\ref{fig:sh2} it appears that the convective heat transport is less efficient compared to the configuration discussed before, in which the matrix was impermeable to solute \cite[for a detailed discussion on the importance of the (im)permeability condition of the solid matrix, see][]{xu2023pore}.
In addition to the effect of the thermal conductivity of the solid, the measurements of \cite{lohsesun2020} refer to relatively high values of porosity.
As predicted by Eq.~\eqref{eq:shraphi}, the larger the porosity, the lower the $\sh$.
The transition from porous convection to unconfined convection is controlled by two physical mechanisms, which are set by the properties of the porous matrix \cite{lohsesun2020}.
On the one hand, the presence of obstacles makes the flow more coherent, with the correlation between temperature fluctuation and vertical velocity enhanced and the counter-gradient convective heat transfer suppressed, leading to heat transfer enhancement.
On the other hand, the convection strength is reduced due the impedance of the obstacle array, leading to heat transfer reduction. 
The variation of $\sh$ with $\raT$ (not $\raM$) is reported in Fig.~\ref{fig:sh3}(c), where the presence of these two distinct regimes is apparent.
For sufficiently large $\raT$ or high porosity, the classical scaling is recovered ($\raT^{1/3}$, \citep{grossmann2000scaling,grossmann2001thermal}).
When the Rayleigh number is lowered, however, the role of the porous structure in confining the flow is critical, and a correlation for the Sherwood number $\sh$ is proposed:
 \begin{equation}
 \sh\approx1+c\phi\left(\frac{H}{\ell_{s}}\right)^{4}\sch^{2}\Rey^{2}(\raT)^{-1},
\label{eq:shrrepr}
\end{equation}
where the Reynolds number $\Rey$ is computed based on the velocity fluctuations and $c=8$ is a fitting parameter.
This scaling is proved to be well approximated by $\sh\sim\raT^{0.65}$ \cite{lohsesun2020}).
The transition between these regimes appears clearly in Fig.~\ref{fig:sh3}(d), when the compensated Sherwood number ($\sh/\raT^{-0.3}$) is shown as a function of the boundary layer thickness [$\delta=H/(2\sh)$] divided by the average pore scale ($l_{s}$).
The situation is schematically illustrated in Fig.~\ref{fig:sh3}(a,b).
When the thickness of the thermal boundary layer is comparable to the averaged pore length scale ($\delta/l_{s}=1$), the transition from one regime to the other occurs. 
In addition to the porous structure and the Rayleigh number, in case of thermal convection, the boundary layer thickness and the heat transfer coefficient are determined also by the value of thermal conductivity of the solid and liquid phases \citep{korba2022effects,zhong2023thermal}.

\section[One-sided convection]{One-sided convection}\label{sec:os}
The one-sided configuration introduced in Sec.~\ref{sec:convconv} is representative of natural instances like geological CO$_2$ sequestration \cite{huppert2014fluid} and mixing in groundwater flows \cite{simmons2001variable}.
In these cases a fluid-saturated porous domain [sketched in Fig.~\ref{fig:comparsion}(c-i)] is allowed to exchange solute through the top boundary.
The system is initially driven by diffusion \cite{depaoli2017solute,slim2014solutal,slim2013dissolution}.
The fluid layer below the upper boundary becomes progressively rich in solute, increasing the density of the liquid phase.
When sufficiently thick, this high-density layer eventually becomes unstable and fingers like structures form \cite{riaz2006onset,slim2010onset}, evolve (i.e., grow and merge) and if the Rayleigh-Darcy number is sufficiently large [$\raM>O(10^3)$] the system may reach a quasi-steady regime.
In this phase the dimensionless solute flux $\widehat{F}$ computed as in Eq.~\eqref{eq:global5b}, indicating the mass of solute dissolved through the top boundary per unit of surface area and time, is nearly constant over time.
For simplicity, hereinafter we will refer to $\widehat{F}$ as the time-averaged value of flux in this constant-flux phase.
The role of the fingers in promoting solute mixing is crucial, as initially proposed by \citep{ennis2005role,xu2006convective}, since the contribution of convection accelerates considerably the dissolution compared to the purely-diffusive case.
The domain progressively saturates with incoming solute, up to the point in which the local concentration difference between the upper fluid layer and the top boundary is reduced, and the dissolution rate suddenly drops. 
This phase is referred to as shutdown regime and it has been accurately described \cite{hewitt2013convective,slim2014solutal,depaoli2017solute,wen2018rayleigh,hewitt2020vigorous}.
A thorough description of the whole dissolution process is provided by \cite{slim2014solutal}.

In this section we will review the results relative to Darcy and pore-scale flows in the one-sided configuration, and we will focus on the dependency of the dissolution rate $\widehat{F}$ on the flow parameters during the constant flux regime.

\subsection{Darcy flows}\label{sec:osdarcy}
When the Darcy model is considered [Eq.~\eqref{eq:cont2}-\eqref{eq:eqadim3}], the flow is uniquely controlled by the Rayleigh-Darcy number $\raM$, similarly to the Rayleigh-B\'enard case discussed in Sec.~\ref{sec:shdarcy}.
Numerical two-dimensional simulations agree on the value of the flux during the constant flux regime, which was initially determined by \cite{hesse2012phd} to be 
\begin{equation}
\widehat{F}=0.017.
\label{eq:osflux2}
\end{equation}
This observation has been later confirmed by a number of numerical studies \cite{pau2010high,elenius2012time,hewitt2013convective,slim2014solutal,depaoli2017solute,wen2018convective,erfani2021dynamics}.
We refer to \cite{erfani2021dynamics} for a review of literature scaling laws in the presence of variations to this problem (anisotropy, geochemistry, etc.).

In the instance of three-dimensional domains, the dynamics is analogue to that discussed above. 
However, due to the large computational costs, only few numerical works are available 
\cite{pau2010high,fu2013pattern,green2018steady}.
A seminal work in the field is presented by \cite{pau2010high}, who performed three-dimensional simulations and estimated the flux to be higher than in the corresponding two-dimensional case~\eqref{eq:osflux2}.
These results refer to $\raM\le9\times10^{3}$, and additional data at larger Rayleigh-Darcy numbers are required to determine at the exact value for $\widehat{F}$, which has been estimated not to exceed 25\% of the two-dimensional case \cite{pau2010high,green2018steady,erfani2021dynamics}.
It is apparent that the additional degree of freedom represented by the third spatial dimension adds significant complexity to the fingering phenomena \cite{pau2010high,fu2013pattern}, with the flow structure being more complex and dynamical \cite{depaoli2022strong}.

\subsection{Pore-scale and Hele-Shaw flows}\label{sec:ospore}
The determination of $\widehat{F}$ has been carried out beyond the Darcy model via pore-scale simulations and experiments, and via Hele-Shaw setups.
As discussed in Sec.~\ref{sec:shdarcy}, a classical argument for Darcy convection requires that $\sh$ scales linearly with $\raM$ \citep{howard_64,malkus_54}. 
The theoretical interpretation is that in natural convection $\sh$ is uniquely controlled by the diffusive boundary layer, and it is independent of the flow interior and any external length scale. 
Only for an exponent of one for $\raM$, i.e., $\sh\sim\raM$, it is possible to have an expression for $\sh$ that is independent of $H$ \citep{howard_64,nield17,amooie2018solutal}.
As a result [see also Eq.~\eqref{eq:global4}], the flux $\widehat{F}$ is expected to be independent of $\raM$, as it emerges from Darcy simulations [e.g., see correlation~\eqref{eq:osflux2}].
Despite this robust theoretical framework, a different scaling for $\sh(\raM)$ was found by many studies.

\begin{figure}[t!]
\center
\includegraphics[width=0.35\textheight]{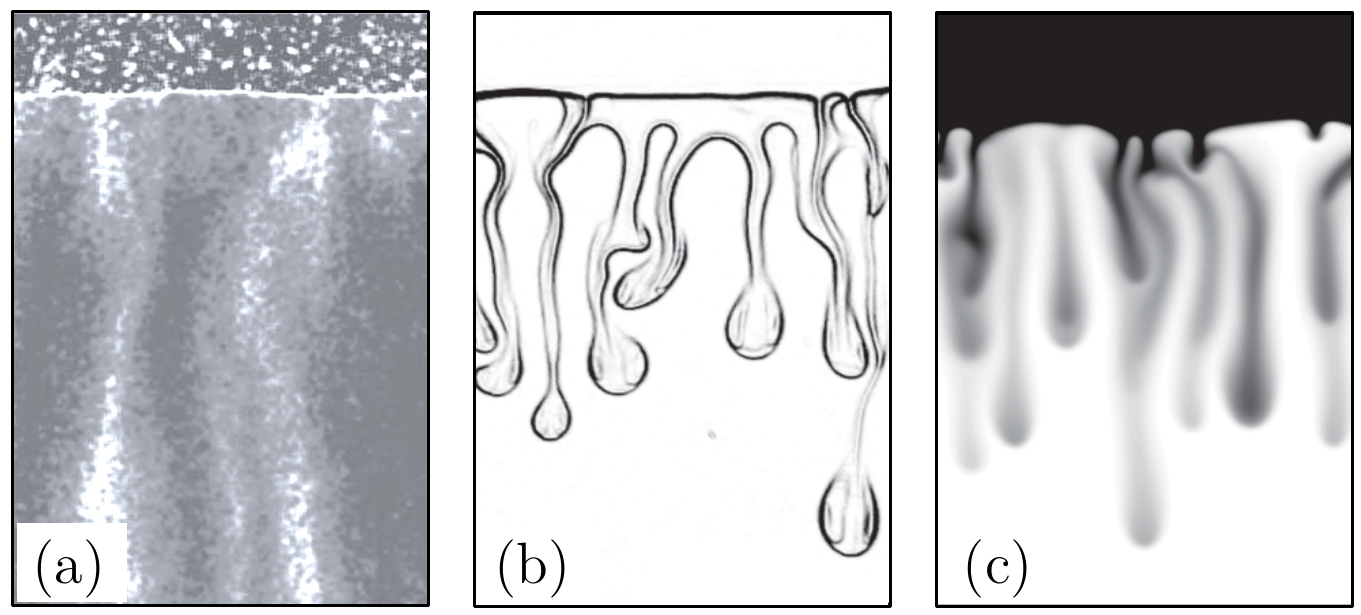}
\caption{\label{fig:sh4}
Examples of one-sided studies. 
(a)~Experiment with MEG in water in bead packs \cite[adapted with permission from][]{neufeld2010convective}. 
(b)~Experiment with propylene-glycol (PG) and water in Hele-Shaw cell \cite[adapted with permission from][]{hidalgo2012scaling}.
(c)~Darcy simulation with non-monotonic density profile \cite[adapted with permission from][]{hidalgo2012scaling}.
}
\end{figure}

Most of the experimental studies investigating one-sided convection have been carried out with the aid of bead packs or Hele-Shaw cells, examples of which are reported in Figs.~\ref{fig:sh4}(a) and \ref{fig:sh4}(b), respectively. 
In the manner, the porous medium consists of a matrix of rigid spheres, typically made of a transparent material to allow optical access to the flow, enclosed in a transparent container.
A Hele-Shaw cell, in turn, is obtained with two parallel and transparent plates separated by a narrow gap $b$ (usually, less than 1~mm thick).
When the fluid velocity in the cell is sufficiently low (gap-based Reynolds number $\ll1$), the flow behaves as a laminar Poiseuille flow, i.e., the gap-average velocity is proportional to the pressure gradient via the inverse of viscosity and to a constant equal to $k=b^{2}/12$, where $k$ is defined as the equivalent permeability of the cell.
Since this formulation represents an analogue of the Darcy law~\eqref{eq:eqadim2}, the Hele-Shaw cell is commonly used as a tool to reproduce a flow through a porous medium. 
Bead packs and Hele-Shaw experiments used to derive scaling laws are discussed in the following.

A first $\sh(\raM)$ scaling was proposed by \cite{neufeld2010convective}, who used experiments in glass beads to mimic one-sided convection in porous media.
The fluids employed were methanol and ethylene-glycol (MEG) and water.
MEG [upper fluid layer in Fig.~\ref{fig:sh4}(a)] is lighter than water when pure, but it presents a non-monotonic density profile as a function of the fraction of water.
As a result, at the interface between the two fluids [identified by the white boundary in Fig.~\ref{fig:sh4}(a)], a heavier mixture forms and originates finger-like instabilities (white structures).
The Sherwood number, estimated by tracking the receding interface between the two fluid layers, was measured to scale as $\sh\sim\raM^{4/5}$, and the result was explained with a phenomenological model based on a boundary layer theory: The lateral solute diffusion from the downward plumes into the upward ones is responsible for the reduction of local concentration gradients and the corresponding density differences driving the flow. 
This translates into a reduction of the flux, making $\sh$ to reduce with respect to the classical scaling.
An analogue approach was employed by \cite{backhaus2011convective}, who used Hele-Shaw cells and a layer of water located vertically above a layer of propylene glycol (PG) [a similar system is shown in Fig.~\ref{fig:sh4}(b)].
They obtained the scaling $\sh\sim\raM^{0.76}$ and identified the plumes spacing as the key parameter controlling the Sherwood number.
Similar results are derived by \citep{tsai2013density} (Hele-Shaw and beads, $\sh\sim\raM^{0.84}$), \cite{ecke2016plume} (Hele-Shaw, $\sh\sim\raM^{0.76}$) and \cite{guo2021novel} (Hele-Shaw, $\sh\sim\raM^{0.95}$).
The discrepancy existing between these sublinear scalings and the linear theoretical \citep{howard_64,malkus_54} and numerical findings in case of Darcy simulations \cite{pau2010high,hewitt2013convective,slim2014solutal,depaoli2017solute,wen2018convective} has been subject of active investigations.

To examine this mismatch, numerical simulations and theoretical arguments were used \cite{hidalgo2012scaling}.
Accurate simulations were employed to mimic the behaviour of the fluids used in the experiments (characterised by a non-monotonic density-concentration curve, and with a concentration-dependent viscosity), which differ from the ones classically considered in Darcy simulations (linear dependency of density with concentration, constant viscosity). 
A snapshot of the concentration field obtained for a Darcy simulation with non-monotonic density profile is shown in Fig.~\ref{fig:sh4}(c).
It was found that the dissolution flux is determined by the mean scalar dissipation rate, $\widehat{\chi}$.
Mixing in porous media has a universal character, and the non-linear behaviour observed needs to be explained with effects not present in the classical Darcy-Boussinesq model.
In particular, the authors observed that several differences exists between this simple Darcy model and the experiments reporting sublinear scalings.
Among the others, they identified three main possible sources of discrepancies: 
(i)~dependency of viscosity with the solute concentration, 
(ii)~non-monotonic behaviour of fluid density with solute concentration, and 
(iii)~compressibility effects (volume change during the process of dissolution).
The conclusion of \cite{hidalgo2012scaling} is that while the concentration-dependent behaviour of viscosity has a minor effect, the role of the non-monotonic density-concentration profiles (shape of the density curves) may considerably affect the Sherwood number scaling law.
The role of some of these fluid properties has been later investigated and will be discussed in the following.
 
The scaling analysis performed by \cite{amooie2018solutal} for non-Boussinesq and compressible flows reveals that the scaling $\sh\approx181.02+0.165\raM$ represents the best fitting for their data.
Therefore, the authors propose that the previously reported sublinear relations could be in part a result of relatively limited parameter range of the simulations (as in the case of \cite{farajzadeh2013empirical}) or in part because the Rayleigh-Darcy number of the experiments lies below the asymptotic limit, i.e., before the classical linear scaling establishes.
 
To avoid a non-monotonic dependency of density with concentration. i.e. to remove the fluid properties as a possible reason of non-linear scaling, experiments in Hele-Shaw cells have been performed.
Potassium permanganate (KMnO$_{4}$) and water are used as analogue fluids, with solid crystals of KMnO$_{4}$ placed on a metal grid located on top of the cell.
Water gradually dissolves the crystals, which remain in a fixed position hold by the mesh, and the resulting interface between the light and the heavy fluid is always fixed and flat.
This methodology, initially introduced by \cite{slim2013dissolution}, allowed to cover a wide range of Rayleigh-Darcy numbers.
In addition, variations of volume and fluid viscosity with solute concentration are negligible.
Results by \cite{ching2017convective} report a linear scaling of $\sh$ with $\raM$.
Later studies \cite{depaoli2020jfm,alipour2020concentration} indicate that within the same value of permeability the scaling $\sh\sim\raM$ holds. 
In general, $\sh$ may still be a function of $\raM$ due to the presence of mechanical dispersion \cite{letelier2019perturbative}.

The works presented indicate that the fluid properties may not be sufficient to justify the non-linear $\sh(\raM)$ scaling observed.
However, other physical mechanisms induced by the Hele-Shaw cell or the dispersion in the porous medium are not present in the classical Darcy model.
These effects, labelled as finite-size effects, maybe be responsible of the non-linear scaling observed, and will be discussed in detail in the Sec.~\ref{sec:osdisp}.

\section[Finite-size effects]{Finite-size effects}\label{sec:osdisp}
Domain features like lateral confinement, thickness-induced Hele-Shaw dispersion and pore-scale dispersion have been identified to play a role on the non-linear scaling of $\sh$ with $\raM$ or the flow structure.
The influence of these finite-size effects on convection will be reviewed in this section.

\subsection{Effect of confinement}\label{sec:confin}
A natural question arising from numerical simulations is what happens when the domain is confined in one of the wall-parallel directions, and we will address this topic here in the frame of Rayeligh-B\'enard, the Rayleigh-Taylor, and the full reservoir-scale flows dynamics.

The flow in a porous Rayleigh-B\'enard system at large $\raM$ consists of two distinct regions (see Sec.~\ref{sec:res2}): (i)~the near-wall region, characterised by the presence of protoplumes, and (ii) the interior of the flow, controlled by megaplumes. 
The average flow structure in each of these regions is quantified by via the time- and horizontally-averaged wavenumber, $k$.
While the near-wall region is hard to be described theoretically, the interior of the flow has been well characterised.
In two dimensions, stability analysis \cite{hewitt2013stability} of the flow interior for $\raM\to\infty$ suggests that $k\sim\raM^{5/14}$, in fair agreement with numerical measurements that give $k\sim\raM^{0.4}$ \cite{hewitt2012ultimate}.
In three dimensions, theoretical results \cite{hewitt2017stability} indicate that $k\sim\raM^{1/2}$, which is in excellent agreement with numerical measurements of \cite{hewitt2014high} and \cite{depaoli2022strong}, who obtained $k\sim\raM^{0.52}$ and  $k\sim\raM^{0.49}$, respectively.
In addition, \cite{pirozzoli2021towards} observed with the aid of numerical simulations that supercells, representing clusters of protoplumes located near the boundaries, are the footprint of the megaplumes dominating the bulk of the flow.
Unexpectedly, the correlation between these flow structures is observed to hold up to very high Rayleigh-Dacry numbers.
This flow structure, however, may be considerably affected by the domain size.

Two-dimensional numerical simulations performed by \cite{wen2015structure} revealed that identifying the wavenumber may be complicated.
Domains with low aspect ratio can dramatically reduce or even suppress convection.
The study shows that the interior structure of a two-dimensional system may result strongly conditioned by the domain width, suggesting that the inter-plume spacing is not unique. 
The authors finally conclude that determining a precise high-$\raM$ scaling of the interior inter-plume spacing will require extremely long simulations in very wide computational domains.

\begin{figure}[t!]
\center
\includegraphics[width=\columnwidth]{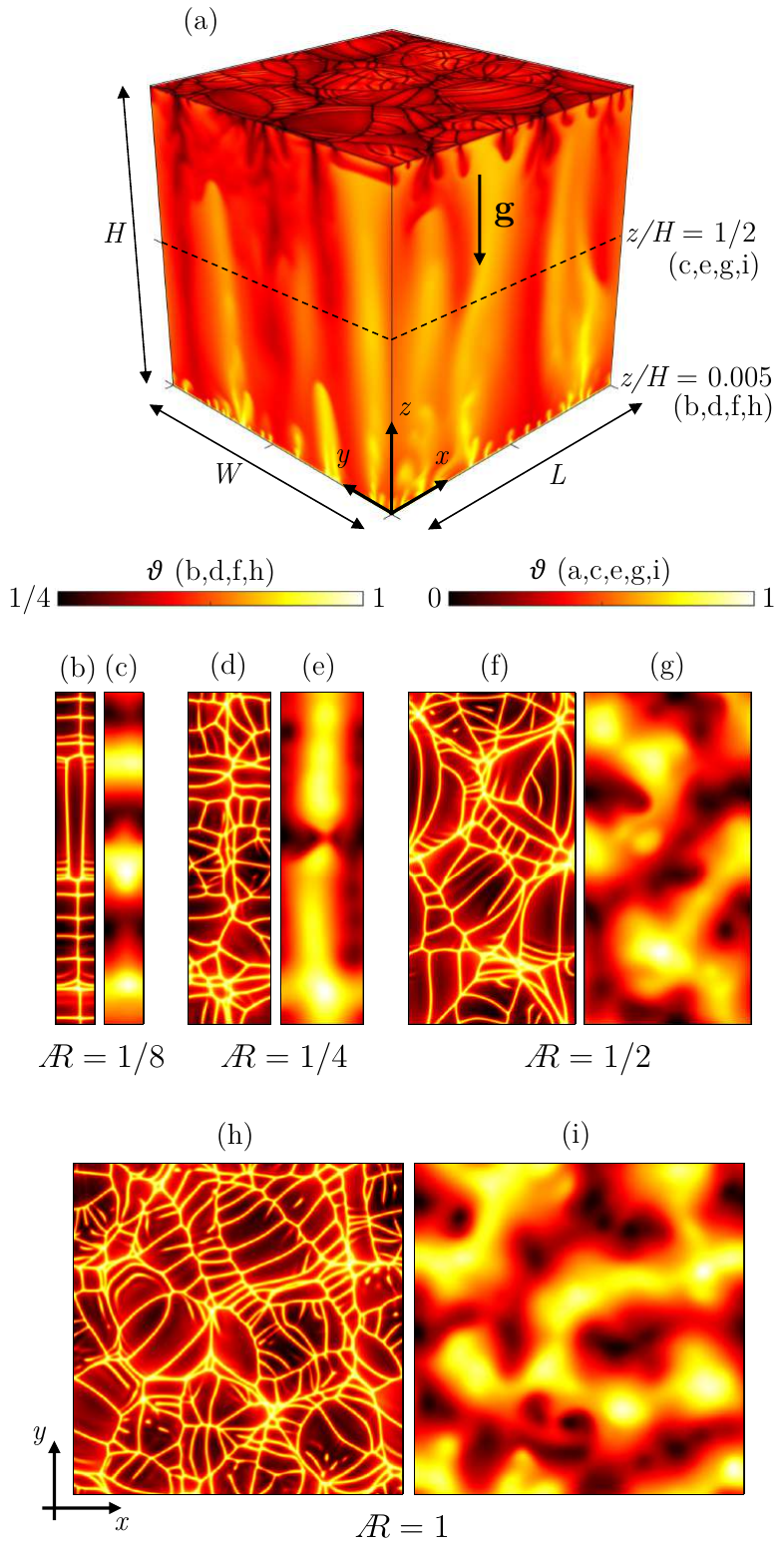}
\caption{\label{fig:dimtr}
Influence of lateral confinement (domain width) on the development of the flow structures.
Three-dimensional Rayleigh-B\'enard simulations performed at $\raM=10^{4}$ are shown \cite[adapted with permission from][]{pirozzoli2021towards,depaoli2022strong}.
(a)~Dimensionless temperature distribution ($\vartheta$) in a cubic domain, being 0 and 1 the values at top and bottom boundaries, respectively, with gravity $\mathbf{g}$ acting along $z$.
Periodic boundary conditions are applied in the wall-parallel directions ($x,y$).
The domain has dimensions $L,W$ and $H$ in directions $x,y$ and $z$, respectively. 
The size of the domain is progressively increased in direction $x$, so that the aspect ratio $\AR=L/H$ increases from $\AR=1/8$ (panels b,c) to $\AR=1$ (panels h,i). 
For each value of $\AR$, temperature fields taken at the centreline $(z=1/2)$ (c,e,g,i) and close to the bottom wall $(z=0.005)$ (b,d,f,h) are shown.
Note that different colorbars apply to centreline and near-wall panels.
}
\end{figure}

In three-dimensions, the effect of the domain confinement has been investigated by \cite{depaoli2022strong}.
They performed numerical simulations at $\raM=10^{4}$ in Rayleigh-B\'enard configuration, in domains having variable extension in one of the wall-parallel directions, namely $x$ in Fig.~\ref{fig:dimtr}(a), and constant extension in the other directions ($W=H$). 
Periodic boundary conditions are applied in the wall-parallel directions.
The relative size of the domain extension in directions $y,z$ with respect to $x$ is quantified by the aspect ratio $\AR=L/H$.
Four values of $\AR$ are considered in Fig.~\ref{fig:dimtr}, with the domain progressively increasing in size from $\AR=1/8$ to $\AR=1$.
The corresponding temperature fields, taken at the centreline $(z=1/2)$ and close to the bottom wall $(z=0.005)$, are shown in Figs.~\ref{fig:dimtr}(b)-(i).
A strong confinement of the domain presents dramatic effects on the flow structures.
For sufficiently large domains, e.g. $\AR=1$, the near-wall cells reported in Fig.~\ref{fig:dimtr}(b) are randomly oriented and show a wide distribution of sizes. 
When the domain width is progressively reduced, the cells are strongly constrained [Figs.~\ref{fig:dimtr}(f,h)] and eventually end up in an extremely ordered pattern [Figs.~\ref{fig:dimtr}(d)]. 
The same applies to the flow structures at the centerline that for small domains ($\AR\le1/4$) form sheet-like plumes.
More quantitative results, estimated by means of the horizontal radial mean wavenumber of these simulations and additional larger domains (not shown here), indicate that the flow structures at the near-wall and in the interior of the flow are strongly constrained by the size of the domain.
They found that at $\raM=10^4$ the flow is independent of the size of the domain for $\AR\ge1$.

Decreasing the size of the computational domain in one direction will inevitably change the flow structure from a three-dimensional towards a two-dimensional character.
This transition has been investigated in the frame of Rayleigh-Taylor instability by \cite{borgnino2021dimensional}.
Among the other indicators, they analysed the evolution of the mixing length, i.e. the time-dependent vertical extension of the tip-to-rear finger distance, to determine whether the system exhibits a two- or three-dimensional behaviour.
They observed that for sufficiently large Rayleigh-Darcy numbers ($\raM>10^{5}$), the growth of the mixing length is always linear in time in two and three dimensions (note that at lower Rayleigh-Darcy numbers the growth of the mixing length may be superlinear \cite{depaoli2019prf,depaoli2022experimental}).
The prefactor of the growth for the mixing length varies, being larger in two dimensions than in three dimensions. 
They performed three-dimensional numerical simulation with triply periodic boundary conditions, in which the dimension of the domain in a direction perpendicular to gravity, defined in the following ``thickness'', is progressively reduced. 
Results indicate that when the thickness diminishes below a certain threshold value, the systems transitions from a three-dimensional to a two-dimensional behaviour.
This critical value corresponds to the wavelength associated with the most unstable mode obtained from linear stability analysis \cite{wooding1962stability,trevelyan2011buoyancy}.
The sharp transition observed in this case is remarkably different than in turbulent convection \cite{boffetta2022dimensional}.
In the turbulent case, the dimensional transition occurs dynamically, i.e. when the width of the mixing region exceeds the confined dimension, and it is smooth due to the co-existence of direct and inverse energy cascades. 

The horizontal domain extension is also a parameter that dramatically affects the evolution of a buoyant current from injection to complete dissolution, e.g. in the configuration sketched in Fig.~\ref{fig:intro}(c) relative to geological sequestration of carbon dioxide.
Using the model for two-phase gravity currents proposed by \cite{macminn2012spreading}, \cite{depaoli2021influence} analyzed the effect of the domain width on the maximum horizontal extension of the current of carbon dioxide. 
They performed two-dimensional simulations in which the domain width is progressively increased, while keeping the domain height and the volume of fluid injected constant. 
It was found that the layer of CO$_2$-rich solution may spread over a horizontal distance greater than 100 times the vertical extension of the layer, indicating that simulations are width-dependent, and very wide domains have to be considered (width to height ratio $\ge140$).

\subsection{Hele-Shaw flows}
\begin{figure}[t!]
\center
\includegraphics[width=0.35\textheight]{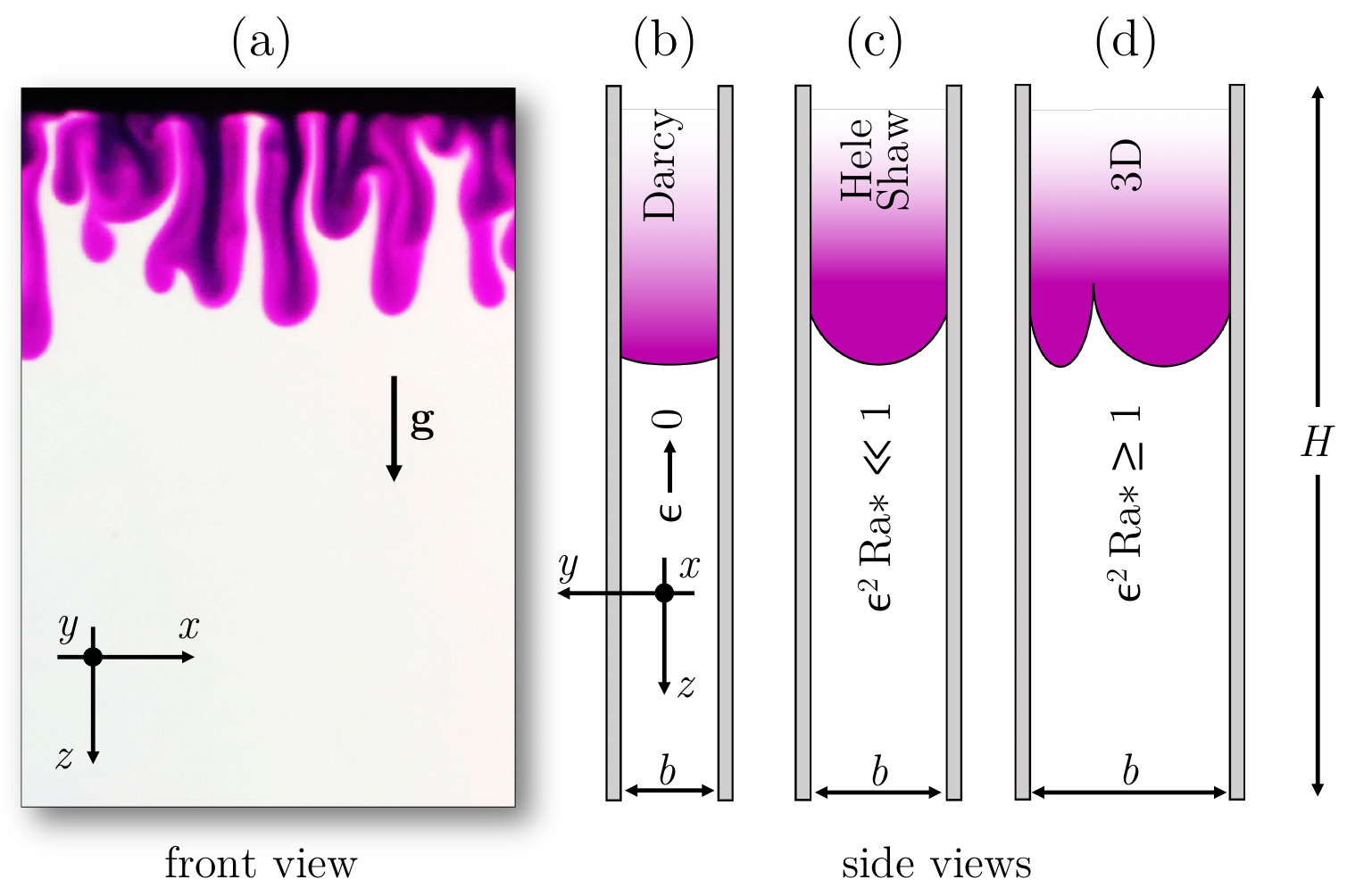}
\caption{\label{fig:sh5}
(a)~Front view of convective flow in a Hele-Shaw cell in one-sided configuration \cite{depaoli2020jfm}, with the solute concentration being constant at top. 
Fluids consists of an aqueous solution of KMnO$_{4}$ (purple to black) and water (white).
The reference frame ($x,y,z$) and the direction along which gravity ($\mathbf{g}$) acts are also indicated.
(b-d)~Schematic representation of the side views of the cell (the thickness $b$ is not to scale with respect to the height $H$).
Three possible flow regimes as identified by \cite{letelier2019perturbative} are shown.
}
\end{figure}

The working principle of the Hele-Shaw apparatus, briefly introduced in Sec.~\ref{sec:ospore}, is illustrated in Fig.~\ref{fig:sh5}(a).
The fluid is contained between two parallel plates separated by a narrow gap of thickness $b$, and the flow obtained in this configuration may be representative of a Darcy flow.
When the flow is dominated by viscous forces (gap-based Reynolds number $\ll1$), the depth-averaged fluid velocity is proportional to the vertical pressure gradient and to the inverse of the viscosity, in analogy to the Darcy law \eqref{eq:eqadim2}.
This proportionality constant, defined as equivalent permeability of the cell, is $k=b^2/12$, and it used to draw a link between Darcy and Hele-Shaw flows.

In convective flows, the driving force of the system is the presence of a solute with concentration $C_0\le C\le C_1$, which produces a maximum density difference $\Delta\rho$ within the domain.
In this frame, the analogy between Hele-Shaw and Darcy flow has been investigated quantitatively by \cite{letelier2019perturbative}, who observed that a combination of fluid properties (Schmidt number, $\sch$), cell geometry (anisotropy ratio, $\epsilon=\sqrt{k}/H$) and flow velocity ($U$, defined in \eqref{eq:eq33}, which depends on $\raM$) determines the flow regime.
They considered an incompressible flow \eqref{eq:cont1}, and averaged the Navier-Stokes and ADE equations, respectively \eqref{eq:ns1} and \eqref{eq:ade1}, in the direction of the gap thickness to obtain the following dimensionless system 
\begin{align}
&\frac{\epsilon^{2}\raM}{\sch}\left[\frac{6}{5}\frac{\partial \mathbf{u^*}}{\partial t^*}+\frac{54}{35}(\mathbf{u^*}\cdot\nabla)\mathbf{u^*}\right]=-\nabla p^* -\mathbf{u^*} +\nonumber\\ 
& + C^* \mathbf{k} 
 +\frac{6}{5}\epsilon^{2} \nabla^{2}\mathbf{u^*}-\frac{2}{35}\epsilon^{2}\raM(\mathbf{u^*}\cdot\nabla C^*)\mathbf{k}
\label{eq:letel1}
\end{align}
\begin{align}
\frac{\partial C^*}{\partial t^*}&+\mathbf{u^*}\cdot\nabla C^*=\frac{1}{\raM}\nabla^{2}C^*+\nonumber\\
+&\frac{2}{35}\epsilon^{2}\raM\nabla\cdot\left [(\mathbf{u^*}\cdot\nabla C^*)\mathbf{u^*}\right]\text{  ,}
\label{eq:letel2}
\end{align}
valid for $\epsilon$ small, $\sch\ge1$ and $\epsilon^{2}\raM\ll1$.
A linear dependency of density with concentration is considered.
In this case $^*$ indicates dimensionless variables where the velocity scale is $U$ defined as \eqref{eq:eq33}, the length scale is $H$, the time scale is $H/U$ and the pressure scale is $\mu U H/k$.
The concentration is made dimensionless as $C^* = (C-C_0)/(C_1-C_0)$ and $\mathbf{k}$ is the unit vector with direction opposite to gravity.
Eq.~\eqref{eq:letel1}-\eqref{eq:letel2} may be respectively interpreted as a Darcy law \eqref{eq:eqadim2} and an advection-diffusion equation \eqref{eq:eqadim3}, both with additional corrective terms taking into account the contribution of inertia and solute redistribution due to the presence of the walls.
In the frame of Hele-Shaw convection, three main regimes have been identified \cite{letelier2019perturbative}:
(i) Darcy regime [Fig.~\ref{fig:sh5}(b)] when $\epsilon\to0$, the concentration profile across the cell gap is nearly uniform and the flow is well described by a Darcy model; 
(ii) Hele-Shaw regime [Fig.~\ref{fig:sh5}(c)] when $\epsilon\ll1$,  $\epsilon^{2}\raM\ll1$ and $\sch\ge1$, characterised a gradient of concentration across the cell gap, but with one single finger; and 
(iii) three-dimensional regime [Fig.~\ref{fig:sh5}(d)], when the parameters do not fall in the above mentioned  limits, the inertial effects become dominant and the fluid layer in the gap is unstable, so that multiple fingers appear across the cell thickness.
It is apparent that the cell geometry plays a key role in determining the flow regime and that all laboratory experiments fall either in the Hele-Shaw regime or in the three-dimensional regime.
With the aid of numerical simulations, \cite{letelier2019perturbative} provided an evidence for the reduction of the scaling exponent (discussed in Sec.\ref{sec:ospore}) for convective flows in the Hele-Shaw regime.

These finding were later confirmed by the laboratory experiments of \cite{depaoli2020jfm}, where the flux has been measured for different values of permeability (i.e., different $b$). 
Note that when the Schmidt number is large \cite[as in the case of][where $\sch= O(10^{3})$]{depaoli2020jfm}, the dispersive effects dominate over to the inertial terms.
As a result, Eq.~\eqref{eq:letel1} reduces to the Darcy law \eqref{eq:eqadim2} with additional dispersive corrections.
These findings suggest within the Hele-Shaw regime the scaling exponent is affected by the anisotropy ratio $\epsilon$, as predicted by \cite{letelier2019perturbative}, possibly explaining the discrepancy observed between Darcy simulations \cite{hidalgo2012scaling} and Hele-Shaw experiments \cite{backhaus2011convective}.

Finally, the theoretical work proposed by \cite{letelier2019perturbative} has been recently generalized by \citep{ulloa2022energetics,letelier2023scaling} to the case of more complex systems characterised by the presence of two layers of fluids with non-monotonic density profiles.
The framework provided in \citep{letelier2023scaling} allows to evaluate and compare the mixing performance of different systems.
They propose a universal law for the evolution of $\sh/\widehat{\chi}$, which is independent of the cell geometry ($\epsilon$) and directly proportional to $\raM$.
Using this theoretical framework, they suggest that a possible reason for the sublinear scaling observed by \cite{backhaus2011convective} is the flow regime (Hele-Shaw regime) in which the experiments are performed.

\subsection{Dispersion in bead packs}\label{sec:dispgran}
Recent developments in experimental techniques allowed accurate and non-invasive measurements of convective dissolution in three-dimensional porous media. 
The studies discussed in Sec.~\ref{sec:ospore} are relative to thin domains, i.e., laboratory experiments in which the dimension of the cell in the direction perpendicular to the transparent walls is much smaller than the other two. 
This confinement may have an effect on the development of the flow structures (see Sec.~\ref{sec:confin}) and on the dissolution efficiency of the system.
We will present here three-dimensional measurements of convection in porous media, and discuss possible approaches to model dispersion in this context.

A remarkable contribution in the field on convection in three-dimensional porous media was presented by \cite{Lister1990}.
This work is original because of the medium used, consisting of a fibrous material, and because of the remarkable visualisations performed.
Beside this work work, most of investigations on convection in three-dimensional porous flows involved the presence of bead packs.
The emergence of tomographic imaging systems over the last years has considerably sped up the research in this field.
In a pioneering work by \cite{shattuck1997convection}, magnetic resonance imaging (MRI) of three-dimensional convective flows in opaque media were presented, and plumes at low Rayleigh-Darcy numbers ($<20\pi^{2}$) were visualized.
Also X-ray computed tomography (CT) imaging scan is now frequently used to study mixing of miscible fluids.
\cite{Wang2016,nakanishi2016experimental} provided correlations for Sherwood as a function of P\'eclet and Rayleigh-Darcy number, and observed a sublinear scaling for $\sh$ with $\raM$, with exponent 0.40 and 0.93 respectively.
The same methodology was employed by \cite{liyanage2019multidimensional}, who reported the emergence of characteristic patterns that closely resemble the dynamical flow structures produced by high-resolution numerical simulations. 
In a later study \cite{liyanage2020direct} the role of viscosity has been also investigated.
While on the one hand \cite{liyanage2019multidimensional,liyanage2020direct} observed that the flow is heavily influenced by dispersion, on the other hand a linear scaling $\sh\sim\raM$ holds, in contrast with previous studies.
This discrepancy may be due to the relatively short range and small values of $\raM$ explored, which is well below the value in correspondence of which the system is observed to attain an asymptotic linear scaling \cite{hewitt2012ultimate,pirozzoli2021towards}.
Employing the same measurement technique but different fluids, \cite{eckel2022spatial} achieved larger Rayleigh-Darcy numbers ($\le 55,000$).
Through qualitative and quantitative observations of flow evolution, they also observed an enhanced longitudinal spreading of the solute, but in this case a sublinear scaling for $Sh(\raM)$ holds.

These works agree upon the fact that dispersion is crucial in determining the $\sh(\raM)$ scaling of the flow, and non-Darcy effects should be included in the models employed \cite{brouzet2022co}. 
Dispersion has been identified as responsible for the early onset of convection \cite{hidalgocarrera2009}. 
In addition, \cite{ghesmat2011effect,erfani2021dynamics} observed that the flow structures are influenced by the strength of dispersion and the dissolution rate $\widehat{F}$ is increased with increasing strength of dispersion. 
However, this finding does not apply in general and it seems to be limited to the range parameters considered \cite{wen2018rayleigh}.
With the aid of laboratory experiments, \cite{liang2018effect} proposed that, in addition to the Rayleigh-Darcy number, a flow with dispersion is controlled by a dispersive Rayleigh-Darcy number 
\begin{equation}
\raD=\frac{UH}{\phi D_{T}}=\frac{\raM D}{D_{T}}, 
\label{eq:dispra}
\end{equation}
with $D_{T}$ the transverse dispersion, $U$ the buoyancy velocity defined in Eq.~\eqref{eq:eq33} and $H$ the domain height.
In geological formations, assigning appropriate values to $D_{T}$ is not trivial, and it has been a debated topic \cite[we refer to][for a thorough review on this subject]{gelhar1992critical}.
The anisotropy ratio $r=D_{L}/D_{T}$ (see Sec.~\ref{sec:disp}) is also important to determine the flow character. 
As a result, the parameter space for convective porous media flows with dispersion is controlled by at least three parameters: $\raM$, $\raD$ and $r$.
In order to quantify the relative importance of molecular diffusion to transverse dispersion, one can introduce the parameter \citep{wen2018rayleigh,tsinober2023numerical} 
\begin{equation}
\Delta =\frac{ \raD}{\raM}=\frac{D}{D_{T}}, 
\label{eq:dispra2}
\end{equation}
that can be used to rewrite the dispersion tensor~\eqref{eq:disp_r0} in dimensionless form as:
\begin{equation}
\frac{\mathbf{D}}{D} = \mathbf{I}+\frac{1}{\Delta U}\left[(r-1)\frac{\mathbf{u}\mathbf{u}}{\vert\mathbf{u}\rvert}+\mathbf{u}\mathbf{I}\right].
    \label{eq:disp_r0b}
\end{equation} 
This expression suggests that the case of pure diffusion is recovered when $D_{T}\ll D$, corresponding to $\Delta\gg1$.

With specific reference to granular media, additional simplifications allow a further characterisation of the flow in the parameter's space.
Considering that the longitudinal dispersivity can be approximated \citep{tsinober2023numerical,depaoli2023convective}
as $\alpha_{L}=D_{L}/U\approx d$, we can rewrite Eq.~\eqref{eq:dispra} as
\begin{equation}
\raD=\frac{UH}{\phi D_{T}}=\frac{UH}{\phi U\alpha_{T}}=\frac{rH}{d}. 
\label{eq:dispra3}
\end{equation}
In bead packs, the permeability can be inferred from the Kozeny-Carman correlation \cite{dullien2012porous,xu2008developing}, i.e. 
\begin{equation}
k=\frac{d^2}{36k_C}\frac{\phi^3}{(1-\phi)^2},
\label{eq:eqck}
\end{equation}
where $k_C=5$ is the Carman constant for monodispersed spheres randomly packed \cite{zaman2010hydraulic}. 
As a result, we can provide an expression for $\raD$ and $\raM$ that is an explicit function of the domain ($g,H$), fluid ($D,\mu,\Delta\rho$) and medium ($\phi,d,r$) properties.
This information is particularly important when we characterise the flow in the three-dimensional  parameters space $(\raD,\raM,r)$, which we will do in the following.

\begin{figure}[t!]
\center
\includegraphics[width=0.35\textheight]{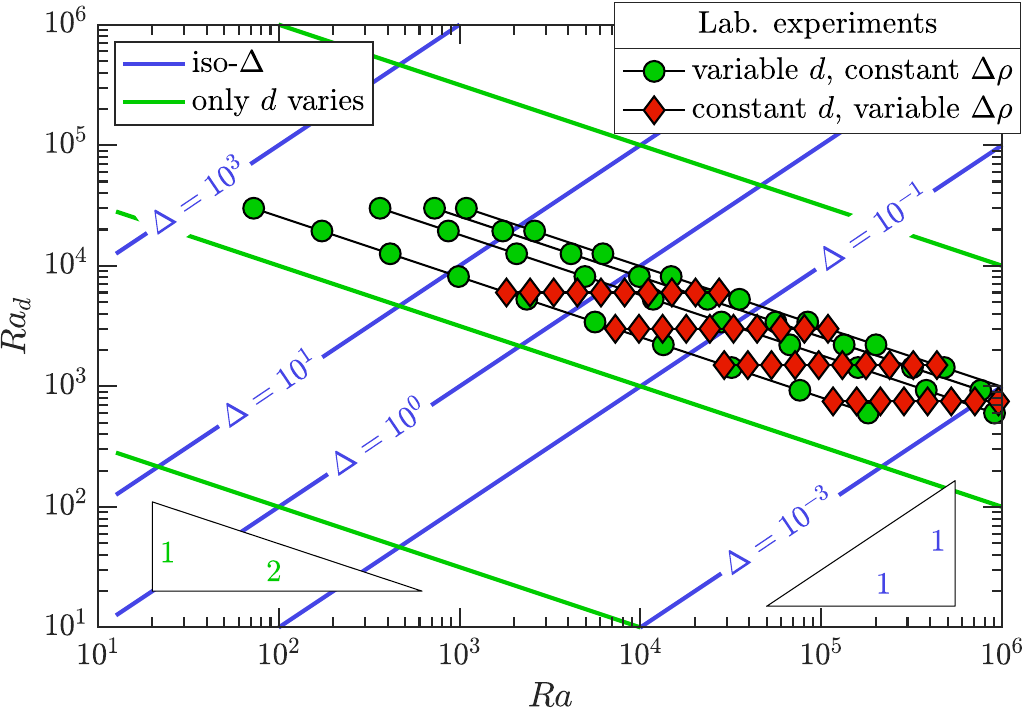}
\caption{\label{fig:shdisp}
Parameters space $(\raM,\raD)$ with indication of iso-$\Delta$ lines (blue lines).
With respect to experiments in bead packs, if only $d$ varies, the parameters $(\raM,\raD)$ are locked to the green curves. 
An example for a realistic range of parameters is shown by symbols (circles), where each line corresponds to one value of density contrast ($\Delta\rho$).
Conversely, if only $\Delta\rho$ is varied in the experiments, $(\raM,\raD)$ are locked to horizontal lines (diamonds).
Triangles indicate the slope of the green and blue lines.
}
\end{figure}

First, we reduce the parameters space to $(\raM,\raD)$ by taking into account \cite{gelhar1992critical} that $r=O(10)$ may be a reasonable approximation for advection dominated systems, and we consider $r=10$. 
Note that no remarkable difference in the flow structure and $\sh$ occurs for $r>10$, provided that $\raM$ and $\raD$ are sufficiently large (namely, $10^{4}$ and $10^{3}$, respectively \citep{wen2018rayleigh}).
For $r\le1$, the flow is qualitatively similar or that observed in the absence of dispersion \cite{hewitt2012ultimate}. 
With respect to the remaining parameters, $\raM$ and $\raD$, we can rewrite both as a function of the beads diameter and find that $\raD\sim 1/d$ and $\raM\sim d^{2}$.
This implies that if we consider an experiment in which only $d$ varies, we are locked to one of the green lines of the parameters space $(\raM,\raD)$ shown in Fig.~\ref{fig:shdisp}, corresponding to $\raD\sim\raM^{-1/2}$.
Using realistic laboratory properties, we obtain that a possible range for the experimental parameters $(\raD,\raM)$ at variable $d$ consists of the circles in Fig.~\ref{fig:shdisp} (each series of circles corresponds to one value of density difference, $\Delta\rho$).
Alternatively, we consider the case in which the medium is fixed ($d$ constant) and the fluid density contrast varies.
Since $\raD$ is independent of any fluid property, a variation of $\Delta\rho$ will correspond to a horizontal line of the parameters space (red symbols in Fig.~\ref{fig:shdisp}, in which each series is a different $d$). 
Finally, we consider the case of a constant value of $\Delta$. 
It follows that this is achieved when $(\Delta\rho)^{-1}d^{-3}$ is constant [blue lines in Fig.~\ref{fig:shdisp}].
This condition is extremely challenging to be obtained experimentally because it implies a simultaneous variation of $\Delta\rho$ and $d$.

\begin{figure}[t!]
\center
\includegraphics[width=0.35\textheight]{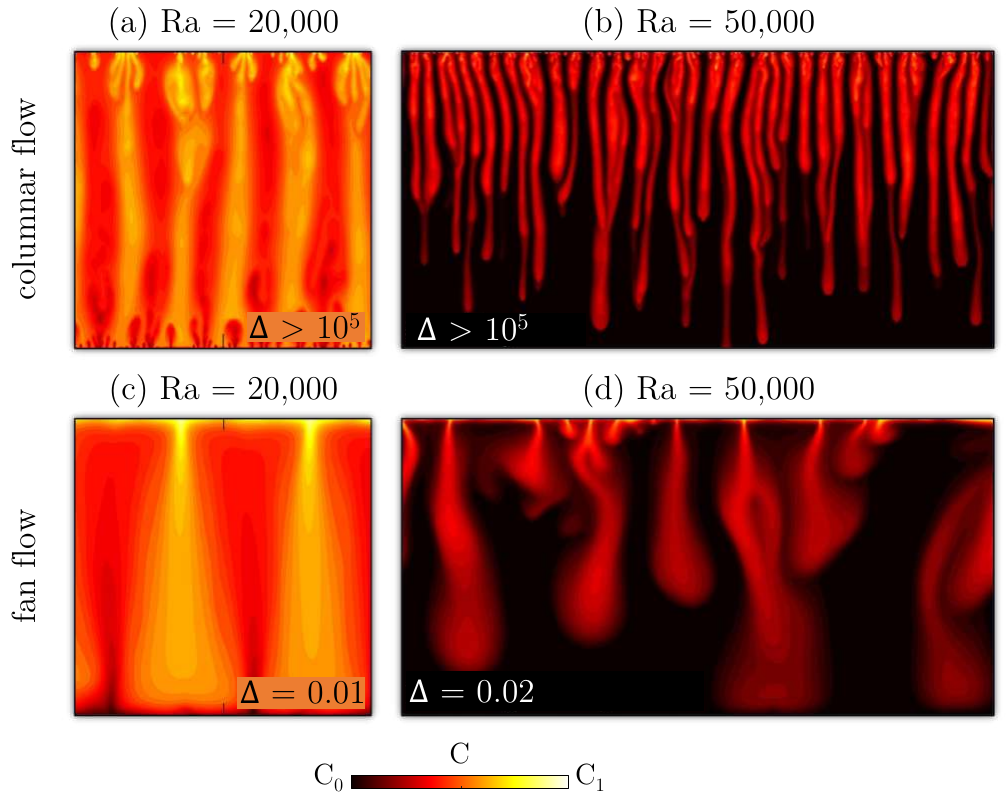}
\caption{\label{fig:shdisp2}
Concentration distribution at high $\raM$ and $r=10$ (Rayleigh-Darcy number and $\Delta$ indicated within each panel).
(a,c)~Rayleigh-B\'enard configuration   
\cite[adapted with permission from][]{wen2018rayleigh}.
(b,d)~One-sided configuration \cite[adapted with permission from][]{liang2018effect}.
When $\Delta\gg1$ (a,b), plumes grow vertically in a symmetric fashion (columnar flow). 
When $\Delta\ll1$ (c,d), dispersion makes the plumes to expand in the horizontal direction (fan flow).
}
\end{figure}

\begin{figure}[t!]
\center
\includegraphics[width=0.35\textheight]{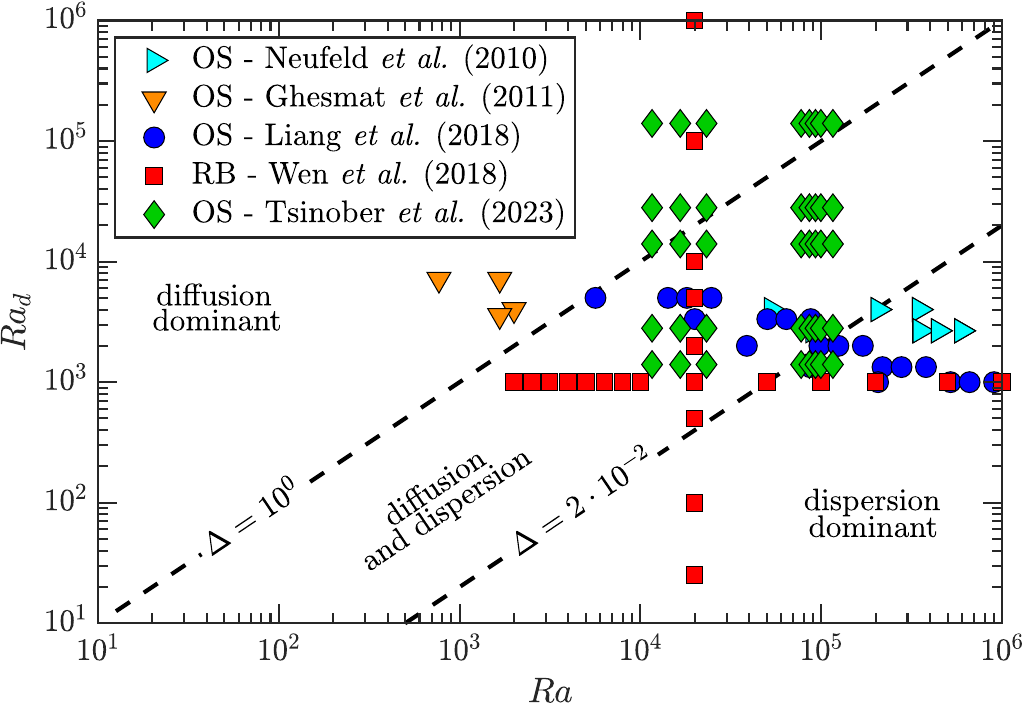}
\caption{\label{fig:shdisp3}
Range of parameters space explored with simulations \cite{ghesmat2011effect,wen2018rayleigh,tsinober2023numerical} and experiments \citep{liang2018effect,neufeld2010convective} with dispersion. 
The configuration (one-sided - OS or Rayleigh-B\'enard - RB) is indicated.
All cases refer to $r=10$, with the exception of \citep{ghesmat2011effect} where also different values of $r$ are considered.
Effect of $r$ on convection is also discussed in \citep{wen2018rayleigh}, but the corresponding data are not reported in this figure.
In this parameters space, $\Delta$ sets the flow behaviour: diffusion dominated [$\Delta<O(1)$], dispersion dominated [$\Delta>0.02$] or influenced by both diffusion and dispersion [$0.02<\Delta<O(1)$].
}
\end{figure}

With the aid of numerical simulations the problem of decoupling two of the governing flow parameters $(\raM,\raD)$ can be solved, and their relative effect on the $\sh$ or $\widehat{F}$ can be investigated.
 \cite{wen2018rayleigh} considered a Rayleigh-B\'enard configuration and investigated systematically a range of flow parameters indicated in Fig.~\ref{fig:shdisp3} (red squares).
The flow structure is mainly ruled by $\Delta$, which determines the mechanism controlling convection.
If $\Delta>10^{5}$, the flow is ruled by molecular diffusion, plumes grow symmetrically [Fig.~\ref{fig:shdisp2}(a)] and the structure is analogue to the symmetric flow observed in the absence of dispersion [see Fig.~\ref{fig:comparsion}(b-i)].
When $\Delta<1$, mechanical dispersion dominates over convection, and its inherent anisotropy ($r\gg1$) sets the non-symmetric flow structure (fan flow) shown in Fig.~\ref{fig:shdisp2}(c), in which plumes widen as they move away from the wall.
A similar behaviour is observed in the corresponding one-sided cases [Fig.~\ref{fig:shdisp2}(b,d)] by \citep{liang2018effect}.

The effect of the flow structure on the Sherwood number was also quantified.
Note that in case of dispersive flows the Sherwood number contains the magnitude of the velocity at the top wall in its definition.
Indeed, while the vertical component of velocity in zero at top (no-penetration), a non-zero velocity parallel to the wall is admitted (free-slip), which produces solute spread due to dispersion.
Alternatively, $\sh$ can be inferred from the time derivative of the total mass of solute in the domain. 
We report in Fig.~\ref{fig:shdisp3} a visual interpretation of the dominant mechanism in each region of the $(\raM,\raD)$ space, where regions controlled by different mechanisms are separated by dashed lines.
\cite{liang2018effect} found that in Rayleigh-B\'enard configuration, when $\Delta>O(1)$ molecular diffusion dominates over mechanical dispersion, although a small contribution of mechanical dispersion may increase $\sh$.
When $0.02<\Delta<O(1)$, both mechanical dispersion and molecular diffusion determine the value of $\sh$.
A linear scaling $\sh\sim\raM$ holds when $\Delta<0.02$, but $\raD$ is also important since it determines the prefactor of the scaling law, as it has been later observed by \cite{erfani2022scaling}.

These results for dispersion-dominated flows ($\Delta<0.02$) could also provide an additional interpretation to the Hele-Shaw experiments of \cite{depaoli2020jfm}, where within one value of cell gap (i.e., one value of permeability and mechanical dispersion), the flux remains nearly constant, i.e. the prefactor is constant.
On the other hand, Hele-Shaw flows do not exhibit transverse dispersion \cite{maes2010experimental}, therefore we believe that such one-to-one comparison between results of dispersion for Hele-Shaw and bead packs flows may not be appropriate.

Recent works investigated the role of mechanical dispersion with the aid of simulations.
A possible complementary approach with respect to the formulation used by \cite{wen2018rayleigh,liang2018effect} is proposed by \cite{dhar2022convective}, where the dispersive Rayleigh number is replaced by a parameter quantifying the strength of longitudinal dispersion compared to molecular diffusion.
More recently, \cite{tsinober2023numerical} performed simulations in one-sided configuration.
They modelled fluids with constant viscosity and linear density-concentration profiles, and derived a linear correlation between $\sh$ and $\raM$, where the prefactor is a function of molecular to dispersive Rayleigh-Darcy numbers, $\raM/\raD$.
This correlation fits well their results, but it does not fully capture the trend predicted by \cite{wen2018rayleigh}.
This difference may possibly be due to several reasons, including the parameters space (and perhaps the different regimes) explored compared to \cite{wen2018rayleigh} as it appears from Fig.~\ref{fig:shdisp3}, where the parameters investigated in some of these studies are reported.
Each of these works involves a specific configuration (Rayleigh-B\'enard or one-sided), a different formulation (different model for the fluids and different dimensionless parameters) and a different region of the parameter space. 
Therefore, providing a precise and general description of convective and dispersive flows in porous media is still not possible, and further studies systematically investigating a broad range of the $(\raM,\raD)$ space are required.

Finally, an novel approach consists of including the effects of dispersion also in the momentum equation.
\cite{gasow2021macroscopic} used two-dimensional pore-scale and Darcy simulations to study a Rayleigh-B\'enard flow.
They observed that the pore-induced dispersion, which may be as strong as buoyancy, affects also the momentum transport and it is determined by two length scales (the pore length scale, proportional to $\sqrt{k}$, and the domain size, $H$).
The authors proposed a two-length-scale diffusion model, in which the pore-scale dispersion is accounted into the momentum transport as a macroscopic diffusion term.
A similar model, which is found to be valid for a wide range of porosity values and is based on the effective viscosity, has been proposed to account for pore-scale effects in advection-dominated systems in the absence of convection \cite{feixiong2020numerical}.

\section{Summary and future perspectives}\label{sec:concl}
In this work, we have reviewed recent developments on convection in porous media. 
We focused on state-of-the-art measurements of dissolution and mixing in archetypal flow configurations.
Despite the well known mathematical formulation of the problem, the role that several physical processes (e.g., finite-size effects) have on the dissolution and mixing is not yet fully understood.
This is also due to the great complexity of the physics involved: convection in porous media is a non-linear phenomenon taking place in multiphase and multiscale systems, eventually located thousands of meters beneath the Earth surface.
Notwithstanding the intrinsic difficulties associated with performing reliable measurements in such systems, remarkable developments have been achieved in recent years. 

The porous Rayleigh-B\'enard configurations, consisting of a fluid-saturated porous slab with fixed density at top and bottom boundary, has been extensively investigated \cite{otero2004high,hewitt2012ultimate,wen2015structure,depaoli2016influence}.
The governing parameter of the flow is the Rayleigh-Darcy number $\raM$, a measure of strength of convection relative to diffusion.
Three-dimensional Darcy simulations performed at unprecedented Rayleigh-Darcy numbers, $O(10^{5})$ have been used, and the existence of new flow features labelled as supercells emerged \cite{pirozzoli2021towards,depaoli2022strong}.
Two-dimensional and three-dimensional simulations have shown that ultimately a linear scaling of the dimensionless dissolution coefficient is attained, namely $\sh\sim\raM$.
While in the two-dimensional case \cite{hewitt2012ultimate,wen2015structure} this scaling sets in at $\raM\le10^{4}$, in three-dimensional flows \cite{pirozzoli2021towards,depaoli2022strong} the ultimate state is expected to take place at $\raM\ge5\times10^{5}$, which is beyond the present numerical capabilities. 
Pore-scale simulations have revealed a more complex scenario, in which the heat/mass transfer is also influenced by porosity \cite{lohsesun2020,gasow2020}, Schmidt number and relative conductivity of fluid and solid phases \cite{korba2022effects,zhong2023thermal}.
These extensive numerical campaigns have led to the development of physics-based correlations for $\sh$ as a function of the flow parameters.
In addition, the relative size of boundary layer and average pore-space has been identified as a critical flow feature controlling pore-scale convection \cite{lohsesun2020}.

The second archetypal configuration considered is the one-sided configuration \cite{hewitt2013convective}, where solute dissolves in an initially solute-free porous domain from the upper boundary, with all other boundaries being impermeable to fluid and solute.
The flow is characterised by an intermediate phase in which the dissolution rate $\widehat{F}$ is quasi steady.
While Darcy simulations report a constant $\raM$-independent value $\widehat{F}$ \cite{hesse2012phd,hewitt2013convective,slim2014solutal,depaoli2017solute}, experiments in bead packs \cite{neufeld2010convective} and Hele-Shaw cells \cite{backhaus2011convective}  revealed that $\widehat{F}$ is a function of $\raM$.
The discrepancy observed has been attributed to non-Darcy effects present in the experiments and not accounted by the simulations \cite{hidalgo2012scaling}. 
This has stimulated further studies focusing on the role of finite-size effects observed in Hele-Shaw \cite{depaoli2020jfm,letelier2019perturbative,ulloa2022energetics,letelier2023scaling} and bead packs experiments \cite{liang2018effect,wen2018rayleigh}. 
The analysis of recent numerical and experimental results \cite{liang2018effect,wen2018rayleigh,tsinober2023numerical} highlights the complexity of this system, which is controlled by at least three parameters, respectively quantifying the relative strength of (i) convection and diffusion ($\raM$), (ii) convection and dispersion ($\raD$), and (iii) longitudinal and transverse dispersion ($r$).
The huge parameters space defined in this way and the need for both numerical and computational studies represents a major challenge in this field.

Improvement of numerical and experimental techniques allowed a detailed characterisation of the flow and a better understating of the phenomena involved. 
The combination of theoretical modelling, numerical simulations and laboratory observations will pave the way to derive and validate large-scale models to be employed in real geophysical and engineering situations.
These findings will be crucial to tackle problems associated with grand societal challenges, such as energy transition and climate change mitigation \cite{dauxois2021confronting}.

To conclude, in Sec.~\ref{sec:methexp} we will briefly review recent advancements in experimental techniques, and in Sec.~\ref{sec:stateart} we will also discuss the importance of additional effects not considered in previous sections of this paper.

\subsection{Recent developments in experimental techniques}\label{sec:methexp}

One intrinsic challenge associated with measurements in porous media consists of the impossibility of optically accessing inner regions of the flow.
An overview of the experimental techniques available to perform measurements in opaque media is presented by \cite{poelma2020measurement}.
Among the different imaging techniques employed for porous media \cite{bear2018modeling}, magnetic resonance imaging (MRI) \cite{suekane2009application,fannir2018two,wang2021quantitative} and X-ray tomography \cite{Wang2016,mahardika2022competition} are the most common, and allow to obtain non-invasive and non-intrusive three-dimensional measurements of inner flow regions.
Despite the advantages mentioned, these techniques are high-priced and typically lack in resolution in both space and time, making fast and small-scale flows hard to measure.
However, thanks to the recent technological progresses, these measurement techniques allowed a detailed characterisation of both medium and flow also at small scales.
Some examples are the X-ray synchrotron microtomography \cite{Hasan2020}, with resolution in space of 3.25~$\mu$m and in time of 6~s.
Recent experiments \cite{do2021x,angulo2022design} have shown that the resolution can be further lowered down to 2.3~$\mu$m, with a technique also allowing for higher resolution in time.
At the time being, similar performances are also achieved by commercial micro-CT systems.
Optical measurement in three-dimensional porous media can be also performed by matching the refractive index of fluid and medium \cite{brouzet2022co,souzy2020velocity,sabbagh2020micro}, provided that a suitable fluid is available.
This is not always granted, since fluids with refractive indexes of interest may come with side effects such as high costs or high hazard \cite{hassan2008flow}.

Additional challenges associated with laboratory experiments, in particular with respect to  geological sequestration of carbon dioxide, consist of reproducing realistic porous media and ambient conditions.
For instance, at the depths at which CO$_{2}$ is supercritical, the pressure is of the order of tens of bars, and performing controlled experiments with optical access is not trivial.
This obstacle has been recently successfully overcome \cite{khosrokhavar2014visualization,croccolo2022}, and the methods proposed may represent a first important step towards investigations in more complex geometries. 
With respect to the design and production of synthetic media at the laboratory scale, microfluidic devices mimicking porous materials are usually made of polydimethylsiloxane (PDMS), which has the drawback of being permeable to CO$_{2}$.
A solution has been recently proposed by \cite{de2022two}, who developed a new method to fabricate a two-dimensional porous medium (regular array of cylinders), consisting of bonding of a patterned photo-lithographed layer on a flat base. 
Additional examples of manufacturing techniques for analogue porous media are provided in \cite{morais2020studying}.
Real geological formations are inherently disordered and heterogeneous, and mimicking this feature in lab models is essential to capture the role of the medium heterogeneities on the solute mixing. 
The technique proposed by \cite{guo2023using} addresses this issue, and it consists of a cell made of 3D-printed elementary blocks designed to be easily re-arranged to obtain a desired permeability field.

Finally, we conclude with an overview of recent developments in experimental techniques employed in Hele-Shaw cells. 
The relative low cost and easy of implementation makes this apparatus widely employed to study buoyancy-driven flows.
Classical optical methods based on light intensity measurements of patterns induced by density (or density gradients) fields, such as Schlieren and related techniques \cite{settles2001schlieren,tropea2007springer}, have been combined or improved to increase the accuracy of the measurements performed. 
Accurate temperature \cite{Letelier2016,noto2023reconstructing} and concentration \cite{slim2013dissolution,ching2017convective,alipour2020concentration} measurement techniques have been recently introduced.
Velocity measurements have been also performed using advanced particle image velocimetry (PIV) and particle tracking velocimetry (PTV) techniques specifically designed for Hele-Shaw flows \cite{ehyaei2014quantitative,kislaya2020psi}, or with the aid of machine learning techniques, namely convolutional neural network (CNN) \cite{kreyenberg2019velocity}.
A separate (i.e., not simultaneous) measurement of scalar and velocity fields complicates the analysis of the phenomena involved and the description the underlying physical mechanisms.
Recently, novel techniques for simultaneous temperature/concentration/velocity measurements have been proposed \cite{anders2020simultaneous,alipour2020concentration}, which are particularly useful to enable reliable comparisons against numerical findings.

\subsection{Additional effects influencing mixing}\label{sec:stateart}
Convection and mixing in real engineering and geophysical problems are far more complex than the idealised conditions depicted in this review, due to the non-ideal medium, fluid, and ambient conditions.
Here we will discuss the influence of conditions not present in the configurations previously discussed, and we will provide some references for the interested readers.

We focused on processes in which the Boussinesq approximation applies, i.e., the density variations induced by the presence of a scalar are only significant within the gravitational term of the momentum (Darcy) equation, and can be neglected elsewhere.
In general, this may not be the case, and a criterion for the applicability of the Boussinesq approximation has been derived \citep{landman2007heat}.
For instance, in case of iso-thermal brine transport, fluid volume changes may be neglected when $\raM\tilde{\rho}_{r}/\Delta\rho\gg1$, being $\Delta\rho$ the maximum density difference and $\tilde{\rho}_{r}$ the reference fluid density.
Interestingly, this condition is independent of $\Delta\rho$ and it is widely fulfilled for geothermal processes, when $\raM\approx10^{1}-10^{3}$ and ${\rho}_{r}/\Delta\rho\approx10^{2}-10^{3}$ \cite{hu2023effects}.
Numerical simulations of the fully-compressible CO$_{2}$ sequestration process suggest that compressibility and non-Boussinesq effects do not significantly impact spreading and mixing \cite{amooie2018solutal}.
An aspect particularly relevant when considering experiments with analogue fluids is that the mixing rate strongly depends on the shape of the fluids density-concentration curve and, in particular, on the position of the maximum of this curve \cite{hidalgo2012scaling}. 
This effect, along with volume variations in the fluid phase \cite{orr2007theory}, may influence the dynamics of the mixing process, and the findings discussed in this review cannot be generalised to fluids with a non-monotonic density-concentration profile or in the presence of significant volume variations.

Geological formations are typically characterised by anisotropic and heterogeneous media. 
The effect of anisotropy has been well characterised by assuming that the permeability tensor is anisotropic \cite{ennis2005onset,cheng2012effect}, and it has been shown that anisotropy is in general favourable since it increases the rate of dissolution and anticipates the onset of convection \cite{emami2017stability,depaoli2016influence,depaoli2017solute}. 
These studies assume that a preferential direction exists, i.e., the permeability tensor takes a diagonal form in a reference frame that usually has a direction aligned with gravity.
This simplified model does not take into account that formations have heterogeneities, which are also source of anisotropy, and discerning these two features of the medium represents a strong simplification.
It has been proposed \cite{green2014steady,green2018steady} that the model of anisotropic medium discussed above (in which the permeability tensor is diagonal in some reference frame) may represent a good candidate to investigate heterogeneous media.
Different models for heterogeneous formations have been introduced, consisting of essentially three categories:
spatially-variable permeability fields \cite{farajzadeh2011effect} (with no preferential direction), long and thin impermeable barriers \cite{elenius2013convective,green2014steady,green2018steady}, and layered formations (i.e., regions in which high- and low-permeability strata alternate) \cite{neufeld2009modelling,hewitt2020buoyancy,soltanian2017dissolution}.
Although a general model for convection in heterogeneous media is not available yet, these studies provide an initial framework to understand the long-term behaviour of these systems.

The role of the fluid properties may also affect the flow evolution and solute mixing. 
The effect of viscosity, for instance, may be crucial in determining the stability of a layer, and we refer to \cite{soboleva2022numerical,kim2023miscible} for a review on this topic.  
Another effect that is increasingly studied is the reactivity of the medium with the fluid: the solute present in the fluid may induce dissolution or precipitation, which corresponds to a variation of the medium porosity and permeability. 
Recently this problem has been actively investigated \cite{erfani2022scaling,hidalgo2015,Fu2015}, also due to the improvement of numerical capabilities.
It has been reported \cite{erfani2021dynamics} that medium morphology modifications occurring in the presence of convective flows affect solute mixing in non-trivial manners.
\cite{cardoso2014geochemistry} showed that the reacting system rock-CO$_{2}$ may be described by a first-order chemical reaction stimulating numerous studies on convective-reactive porous media flows, reviewed in \citep{de2016chemo,de2020chemo}.

Finally, the effect of the ambient flow conditions may be also important \cite{emami2015co2}. 
It was observed that the presence of a background flow influences the onset of convection \cite{hassanzadeh2009effect}. 
Experiments in one-sided configuration \cite{michel2020role} revealed that while convection may be hindered and suppressed, dispersion enhances, with an overall  contribution with respect to flux in the absence of background flow that can be positive, negative or neutral. 
\cite{tsinober2022role} observed with the aid of simulations that three regimes exist, in which convection dominates, background flow dominates, or these two contributions have the same strength.
These results are relevant, since they can contribute to derive new models suitable for prediction of dissolution at the scale of the reservoir \cite{macminn2012spreading,depaoli2021influence,szulczewski2013carbon} and through the entire lifetime of a buoyant current in a porous formation \cite{wang2022analysis,macminn2013buoyant,szulczewski2013carbon,hidalgo2013dynamics}. 

\backmatter

\bmhead{Acknowledgments}
Hugo Ulloa and Diego Perissutti are gratefully acknowledged for their feedback on the early draft of this manuscript. 
The Referees are gratefully acknowledged for their constructive comments.
Duncan Hewitt, Linfeng Jiang, Shuang Liu and Yan Jin are also acknowledged for providing some of the data presented in this work.
This research was funded in part by the Austrian Science Fund (FWF) [Grant J-4612].
The author acknowledges the TU Wien University Library for financial support through its Open Access Funding Program.
This project has received funding from the European Union's Horizon Europe research and innovation programme under the Marie Sk\l odowska-Curie grant agreement No.~101062123.

\section*{Declarations}

\bmhead{Data availability statement} The data supporting the findings of this study are available within the article and any other data can be made available on reasonable request.

\bmhead{Open Access}
This article is licensed under a Creative Commons Attribution 4.0 International License, which permits use, sharing, adaptation, distribution and reproduction in any medium or format, as long as you give appropriate credit to the original author(s) and the source, provide a link to the Creative Commons licence, and indicate if changes were made. 
The images or other third party material in this article are included in the article's Creative Commons licence, unless indicated otherwise in a credit line to the material. If material is not included in the article's Creative Commons licence and your intended use is not permitted by statutory regulation or exceeds the permitted use, you will need to obtain permission directly from the copyright holder. To view a copy of this licence, visit \href{http://creativecommons.org/licenses/by/4.0/}{http://creativecommons.org/licenses/by/4.0/}.

\bibliography{bibliography}

\end{document}